\documentclass[aps,prd,preprint,superscriptaddress,showpacs,preprintnumbers,amsmath,amssymb]{revtex4-1}

\usepackage[dvipdfmx]{graphicx}
\usepackage{dcolumn}
\usepackage{bm}
\usepackage{subfigure}
\usepackage{multirow}
\usepackage{epstopdf}
\usepackage{color}

\begin{document}

\preprint{\vbox{ \hbox{   }
    \hbox{BELLE-CONF-1608}
}}

\title{ \quad\\[0.5cm]  Measurement of the $\tau$ lepton polarization in the decay ${\bar B} \rightarrow D^* \tau^- {\bar \nu_{\tau}}$}

\date{\today}

\noaffiliation
\affiliation{Aligarh Muslim University, Aligarh 202002}
\affiliation{University of the Basque Country UPV/EHU, 48080 Bilbao}
\affiliation{Beihang University, Beijing 100191}
\affiliation{University of Bonn, 53115 Bonn}
\affiliation{Budker Institute of Nuclear Physics SB RAS, Novosibirsk 630090}
\affiliation{Faculty of Mathematics and Physics, Charles University, 121 16 Prague}
\affiliation{Chiba University, Chiba 263-8522}
\affiliation{Chonnam National University, Kwangju 660-701}
\affiliation{University of Cincinnati, Cincinnati, Ohio 45221}
\affiliation{Deutsches Elektronen--Synchrotron, 22607 Hamburg}
\affiliation{University of Florida, Gainesville, Florida 32611}
\affiliation{Department of Physics, Fu Jen Catholic University, Taipei 24205}
\affiliation{Justus-Liebig-Universit\"at Gie\ss{}en, 35392 Gie\ss{}en}
\affiliation{Gifu University, Gifu 501-1193}
\affiliation{II. Physikalisches Institut, Georg-August-Universit\"at G\"ottingen, 37073 G\"ottingen}
\affiliation{SOKENDAI (The Graduate University for Advanced Studies), Hayama 240-0193}
\affiliation{Gyeongsang National University, Chinju 660-701}
\affiliation{Hanyang University, Seoul 133-791}
\affiliation{University of Hawaii, Honolulu, Hawaii 96822}
\affiliation{High Energy Accelerator Research Organization (KEK), Tsukuba 305-0801}
\affiliation{J-PARC Branch, KEK Theory Center, High Energy Accelerator Research Organization (KEK), Tsukuba 305-0801}
\affiliation{Hiroshima Institute of Technology, Hiroshima 731-5193}
\affiliation{IKERBASQUE, Basque Foundation for Science, 48013 Bilbao}
\affiliation{University of Illinois at Urbana-Champaign, Urbana, Illinois 61801}
\affiliation{Indian Institute of Science Education and Research Mohali, SAS Nagar, 140306}
\affiliation{Indian Institute of Technology Bhubaneswar, Satya Nagar 751007}
\affiliation{Indian Institute of Technology Guwahati, Assam 781039}
\affiliation{Indian Institute of Technology Madras, Chennai 600036}
\affiliation{Indiana University, Bloomington, Indiana 47408}
\affiliation{Institute of High Energy Physics, Chinese Academy of Sciences, Beijing 100049}
\affiliation{Institute of High Energy Physics, Vienna 1050}
\affiliation{Institute for High Energy Physics, Protvino 142281}
\affiliation{Institute of Mathematical Sciences, Chennai 600113}
\affiliation{INFN - Sezione di Torino, 10125 Torino}
\affiliation{Advanced Science Research Center, Japan Atomic Energy Agency, Naka 319-1195}
\affiliation{J. Stefan Institute, 1000 Ljubljana}
\affiliation{Kanagawa University, Yokohama 221-8686}
\affiliation{Institut f\"ur Experimentelle Kernphysik, Karlsruher Institut f\"ur Technologie, 76131 Karlsruhe}
\affiliation{Kavli Institute for the Physics and Mathematics of the Universe (WPI), University of Tokyo, Kashiwa 277-8583}
\affiliation{Kennesaw State University, Kennesaw, Georgia 30144}
\affiliation{King Abdulaziz City for Science and Technology, Riyadh 11442}
\affiliation{Department of Physics, Faculty of Science, King Abdulaziz University, Jeddah 21589}
\affiliation{Korea Institute of Science and Technology Information, Daejeon 305-806}
\affiliation{Korea University, Seoul 136-713}
\affiliation{Kyoto University, Kyoto 606-8502}
\affiliation{Kyungpook National University, Daegu 702-701}
\affiliation{\'Ecole Polytechnique F\'ed\'erale de Lausanne (EPFL), Lausanne 1015}
\affiliation{P.N. Lebedev Physical Institute of the Russian Academy of Sciences, Moscow 119991}
\affiliation{Faculty of Mathematics and Physics, University of Ljubljana, 1000 Ljubljana}
\affiliation{Ludwig Maximilians University, 80539 Munich}
\affiliation{Luther College, Decorah, Iowa 52101}
\affiliation{University of Maribor, 2000 Maribor}
\affiliation{Max-Planck-Institut f\"ur Physik, 80805 M\"unchen}
\affiliation{School of Physics, University of Melbourne, Victoria 3010}
\affiliation{Middle East Technical University, 06531 Ankara}
\affiliation{University of Miyazaki, Miyazaki 889-2192}
\affiliation{Moscow Physical Engineering Institute, Moscow 115409}
\affiliation{Moscow Institute of Physics and Technology, Moscow Region 141700}
\affiliation{Graduate School of Science, Nagoya University, Nagoya 464-8602}
\affiliation{Kobayashi-Maskawa Institute, Nagoya University, Nagoya 464-8602}
\affiliation{Nara University of Education, Nara 630-8528}
\affiliation{Nara Women's University, Nara 630-8506}
\affiliation{National Central University, Chung-li 32054}
\affiliation{National United University, Miao Li 36003}
\affiliation{Department of Physics, National Taiwan University, Taipei 10617}
\affiliation{H. Niewodniczanski Institute of Nuclear Physics, Krakow 31-342}
\affiliation{Nippon Dental University, Niigata 951-8580}
\affiliation{Niigata University, Niigata 950-2181}
\affiliation{University of Nova Gorica, 5000 Nova Gorica}
\affiliation{Novosibirsk State University, Novosibirsk 630090}
\affiliation{Osaka City University, Osaka 558-8585}
\affiliation{Osaka University, Osaka 565-0871}
\affiliation{Pacific Northwest National Laboratory, Richland, Washington 99352}
\affiliation{Panjab University, Chandigarh 160014}
\affiliation{Peking University, Beijing 100871}
\affiliation{University of Pittsburgh, Pittsburgh, Pennsylvania 15260}
\affiliation{Punjab Agricultural University, Ludhiana 141004}
\affiliation{Research Center for Electron Photon Science, Tohoku University, Sendai 980-8578}
\affiliation{Research Center for Nuclear Physics, Osaka University, Osaka 567-0047}
\affiliation{Theoretical Research Division, Nishina Center, RIKEN, Saitama 351-0198}
\affiliation{RIKEN BNL Research Center, Upton, New York 11973}
\affiliation{Saga University, Saga 840-8502}
\affiliation{University of Science and Technology of China, Hefei 230026}
\affiliation{Seoul National University, Seoul 151-742}
\affiliation{Shinshu University, Nagano 390-8621}
\affiliation{Showa Pharmaceutical University, Tokyo 194-8543}
\affiliation{Soongsil University, Seoul 156-743}
\affiliation{University of South Carolina, Columbia, South Carolina 29208}
\affiliation{Stefan Meyer Institute for Subatomic Physics, Vienna 1090}
\affiliation{Sungkyunkwan University, Suwon 440-746}
\affiliation{School of Physics, University of Sydney, New South Wales 2006}
\affiliation{Department of Physics, Faculty of Science, University of Tabuk, Tabuk 71451}
\affiliation{Tata Institute of Fundamental Research, Mumbai 400005}
\affiliation{Excellence Cluster Universe, Technische Universit\"at M\"unchen, 85748 Garching}
\affiliation{Department of Physics, Technische Universit\"at M\"unchen, 85748 Garching}
\affiliation{Toho University, Funabashi 274-8510}
\affiliation{Tohoku Gakuin University, Tagajo 985-8537}
\affiliation{Department of Physics, Tohoku University, Sendai 980-8578}
\affiliation{Earthquake Research Institute, University of Tokyo, Tokyo 113-0032}
\affiliation{Department of Physics, University of Tokyo, Tokyo 113-0033}
\affiliation{Tokyo Institute of Technology, Tokyo 152-8550}
\affiliation{Tokyo Metropolitan University, Tokyo 192-0397}
\affiliation{Tokyo University of Agriculture and Technology, Tokyo 184-8588}
\affiliation{University of Torino, 10124 Torino}
\affiliation{Toyama National College of Maritime Technology, Toyama 933-0293}
\affiliation{Utkal University, Bhubaneswar 751004}
\affiliation{Virginia Polytechnic Institute and State University, Blacksburg, Virginia 24061}
\affiliation{Wayne State University, Detroit, Michigan 48202}
\affiliation{Yamagata University, Yamagata 990-8560}
\affiliation{Yonsei University, Seoul 120-749}
  \author{A.~Abdesselam}\affiliation{Department of Physics, Faculty of Science, University of Tabuk, Tabuk 71451}
  \author{I.~Adachi}\affiliation{High Energy Accelerator Research Organization (KEK), Tsukuba 305-0801}\affiliation{SOKENDAI (The Graduate University for Advanced Studies), Hayama 240-0193}
  \author{K.~Adamczyk}\affiliation{H. Niewodniczanski Institute of Nuclear Physics, Krakow 31-342}
  \author{H.~Aihara}\affiliation{Department of Physics, University of Tokyo, Tokyo 113-0033}
  \author{S.~Al~Said}\affiliation{Department of Physics, Faculty of Science, University of Tabuk, Tabuk 71451}\affiliation{Department of Physics, Faculty of Science, King Abdulaziz University, Jeddah 21589}
  \author{K.~Arinstein}\affiliation{Budker Institute of Nuclear Physics SB RAS, Novosibirsk 630090}\affiliation{Novosibirsk State University, Novosibirsk 630090}
  \author{Y.~Arita}\affiliation{Graduate School of Science, Nagoya University, Nagoya 464-8602}
  \author{D.~M.~Asner}\affiliation{Pacific Northwest National Laboratory, Richland, Washington 99352}
  \author{T.~Aso}\affiliation{Toyama National College of Maritime Technology, Toyama 933-0293}
  \author{H.~Atmacan}\affiliation{Middle East Technical University, 06531 Ankara}
% ...
\collaboration{The Belle Collaboration}

\noaffiliation

\begin{abstract}

We report the first measurement of the $\tau$ lepton polarization in the decay ${\bar B} \rightarrow D^* \tau^- {\bar\nu_{\tau}}$ as well as a new measurement of the ratio of the branching fractions $R(D^{*}) = \mathcal{B}({\bar B} \rightarrow D^* \tau^- {\bar\nu_{\tau}}) / \mathcal{B}({\bar B} \rightarrow D^* \ell^- {\bar\nu_{\ell}})$, where $\ell^-$ denotes an electron or a muon, with the decays $\tau^- \rightarrow \pi^- \nu_{\tau}$ and $\tau^- \rightarrow \rho^- \nu_{\tau}$. We use the full data sample of $772 \times 10^6$ $B{\bar B}$ pairs accumulated with the Belle detector at the KEKB electron-positron collider. Our preliminary results, $R(D^*) = 0.276 \pm 0.034{\rm (stat.)} ^{+0.029} _{-0.026}{\rm (syst.)}$ and $P_{\tau} = -0.44 \pm 0.47 {\rm (stat.)} ^{+0.20} _{-0.17} {\rm (syst.)}$, are consistent with the theoretical predictions of the Standard Model within $0.6$ standard deviation.

\end{abstract}

\maketitle

{\renewcommand{\thefootnote}{\fnsymbol{footnote}}}
\setcounter{footnote}{0}

\section{Introduction}

Semitauonic $B$ meson decays with $b \rightarrow c \tau^- {\bar\nu_{\tau}}$~\cite{cite-CC} transitions are sensitive to new physics (NP) beyond the standard model (SM) involving non-universal coupling to heavy fermions. One prominent candidate for NP is the Two Higgs Doublet Model (2HDM)~\cite{cite-2HDM}, which has an additional Higgs doublet and therefore introduces two neutral and two charged Higgs bosons in addition to the SM Higgs boson. The charged Higgs bosons may contribute to the $b \rightarrow c \tau^- {\bar \nu_{\tau}}$ process, modifying its branching fraction and decay kinematics.

Exclusive semitauonic decays of the type ${\bar B} \rightarrow D^{(*)} \tau^- {\bar \nu_{\tau}}$ have been studied by Belle~\cite{cite-Belle1,cite-Belle2,cite-Belle3,cite-Belle4}, BaBar~\cite{cite-BaBar1,cite-BaBar2} and LHCb~\cite{cite-LHCb}. The experiments typically measure the ratios of branching fractions,
\begin{eqnarray}
  R(D^{(*)}) \equiv \frac{\mathcal{B}({\bar B} \rightarrow D^{(*)} \tau^- {\bar\nu_{\tau}})}{\mathcal{B}({\bar B} \rightarrow D^{(*)} \ell^- {\bar\nu_{\ell}})}\label{eq-rdstar}
\end{eqnarray}
where the denominator is the average for $\ell^- \in \{e^-, \mu^-\}$. The ratio cancels uncertainties common to the numerator and the denominator. These include the Cabibbo-Kobayashi-Maskawa matrix element $|V_{cb}|$ and many of the theoretical uncertainties on hadronic form factors and experimental reconstruction effects. The current averages of the three experiments~\cite{cite-Belle3,cite-Belle4,cite-BaBar2,cite-LHCb} are $R(D) = 0.397 \pm 0.040 \pm 0.028$ and $R(D^*) = 0.316 \pm 0.016 \pm 0.010$, which are within $1.9\sigma$ and $3.3\sigma$~\cite{cite-HFAG} of the SM predictions of $R(D) = 0.299 \pm 0.011$~\cite{cite-RD2} or $0.300 \pm 0.008$~\cite{cite-RD} and $R(D^*) = 0.252 \pm 0.003$~\cite{cite-RDst}, respectively. Here, $\sigma$ represents the standard deviation.

In addition to $R(D^{(*)})$, the polarization of the $\tau$ lepton and the $D^*$ meson is also sensitive to NP~\cite{cite-TW_Dtaunu,cite-TW}. The polarization of the $\tau$ lepton ($P_{\tau}$) is defined by
\begin{eqnarray}
  P_{\tau} &=& \frac{\Gamma^+ - \Gamma^-}{\Gamma^+ + \Gamma^-},
\end{eqnarray}
where $\Gamma^{\pm}$ denotes the decay rate of ${\bar B} \rightarrow D^{(*)} \tau^- {\bar \nu_{\tau}}$ with a $\tau$ helicity of $\pm 1/2$. The SM predicts $P_{\tau} = 0.325 \pm 0.009$ for ${\bar B} \rightarrow D \tau^- {\bar \nu_{\tau}}$~\cite{cite-TW_Dtaunu} and $P_{\tau} = -0.497 \pm 0.013$ for ${\bar B} \rightarrow D^* \tau^- {\bar \nu_{\tau}}$~\cite{cite-TW,cite-TWpol}. The $\tau$ polarization is accessible in two-body hadronic $\tau$ decays with the following formulae~\cite{cite-Taupol}:
\begin{eqnarray}
  \frac{1}{\Gamma} \frac{d\Gamma}{d\cos\theta_{\rm hel}} = \frac{1}{2} (1 + \alpha P_{\tau} \cos\theta_{\rm hel}),\label{eq-coshel}\\
  \alpha =
  \begin{cases}
    1 & \text{for pseudo-scalar mesons}\\
    \frac{m_{\tau}^2 - 2m_V^2}{m_{\tau}^2 + 2m_V^2} & \text{for vector mesons,}\label{eq-alpha}
  \end{cases}
\end{eqnarray}
where $\Gamma$, $m_{\tau}$ and $m_V$ are, respectively, the decay rate of ${\bar B} \rightarrow D^{(*)} \tau^- {\bar\nu_{\tau}}$ and the masses of the $\tau$ lepton and the vector meson from the $\tau$ decay. The helicity angle, $\theta_{\rm hel}$, is the opening angle between the momentum vectors of the virtual $W$ boson and of the $\tau$-daughter meson in the rest frame of the $\tau$. The parameter $\alpha$ describes the sensitivity to  $P_{\tau}$ for each $\tau$-decay mode; in particular, $\alpha = 0.45$ for the decay $\tau^- \rightarrow \rho^- \nu_{\tau}$.

In this paper, we report a new measurement of $R(D^*)$ in the hadronic $\tau$ decay modes $\tau^- \rightarrow \pi^- \nu_{\tau}$ and $\rho^- \nu_{\tau}$. This measurement is statistically independent of the previous Belle measurements~\cite{cite-Belle3,cite-Belle4}, with a different background composition. We also report the first measurement of $P_{\tau}$ for the decay ${\bar B} \rightarrow D^* \tau^- {\bar\nu_{\tau}}$.

\section{Detector and Data Samples}

We use the full $\Upsilon(4S)$ data sample containing $772 \times 10^6 B\bar{B}$ pairs recorded with the Belle detector~\cite{cite-Belle} at the asymmetric-beam-energy $e^+ e^-$ collider KEKB~\cite{cite-KEKB}. The Belle detector is a large-solid-angle magnetic spectrometer that consists of a silicon vertex detector (SVD), a 50-layer central drift chamber (CDC), an array of aerogel threshold Cherenkov counters (ACC), a barrel-like arrangement of time-of-flight scintillation counters (TOF), and an electromagnetic calorimeter (ECL) comprised of CsI(Tl) crystals located inside a superconducting solenoid coil that provides a 1.5~T magnetic field.  An iron flux-return located outside of the coil is instrumented to detect $K_L^0$ mesons and to identify muons (KLM). The detector is described in detail elsewhere~\cite{cite-Belle}. Two inner detector configurations were used. A 2.0 cm radius beampipe and a 3-layer silicon vertex detector was used for the first sample of $152 \times 10^6 B\bar{B}$ pairs, while a 1.5 cm radius beampipe, a 4-layer silicon detector and a small-cell inner drift chamber were used to record the remaining $620 \times 10^6 B\bar{B}$ pairs~\cite{cite-SVD2}.  

The signal selection criteria and the signal and background probability density functions (PDFs) used in this measurement rely on the use of Monte Carlo (MC) simulation samples. These samples are generated by the software packages EvtGen~\cite{cite-EvtGen} and PYTHIA~\cite{cite-PYTHIA}; final-state radiation is generated by PHOTOS~\cite{cite-PHOTOS}. Detector responses are fully simulated with the Belle detector simulator based on GEANT3~\cite{cite-GEANT}.

The signal decay ${\bar B} \rightarrow D^* \tau^- {\bar\nu_{\tau}}$ (signal mode) is generated with a decay model based on the heavy quark effective theory (HQET)~\cite{cite-TW}. We use the current world-average values for the form-factor parameters $\rho^2 = 1.207 \pm 0.015 \pm 0.021$, $R_1 = 1.403 \pm 0.033$ and $R_2 = 0.854 \pm 0.020$~\cite{cite-HFAG}, which are based on the parameterization in Ref.~\cite{cite-HQETFF}. Decays of the type ${\bar B} \rightarrow D^* \ell^- {\bar\nu_{\ell}}$ (normalization mode), which are used for the denominator of $R(D^*)$, are also modeled with HQET using the above form-factor values. Background from semileptonic decays to orbitally-excited charmed mesons ${\bar B} \rightarrow D^{**} \ell^- {\bar\nu_{\ell}}$, where $D^{**}$ denotes $D_1$, $D_2^*$, $D'_1$ or $D_0^*$, are generated with the ISGW model~\cite{cite-ISGW} with their kinematic distributions reweighted to match the dynamics predicted by the LLSW model~\cite{cite-LLSW}. Additionally, theoretically-predicted radial excitation states $D^{(*)}(2S)$ are assumed to fill the gap between the inclusively-measured and the sum of the exclusively-measured branching fractions of ${\bar B} \rightarrow X_c \ell^- {\bar \nu_{\ell}}$~\cite{cite-Dssln}. The MC sample of ${\bar B} \rightarrow D^{**} \tau^- {\bar\nu_{\tau}}$ is produced with the ISGW model. The branching fractions are assigned according to their theoretical estimates~\cite{cite-Dsstn}. The remaining background MC samples are comprised of mostly hadronic $B$ meson decays and light quark production processes ($q = u, d, s, c$). The sample sizes of the signal, ${\bar B} \rightarrow D^{**} \ell^- {\bar \nu_{\ell}}$, ${\bar B} \rightarrow D^{**} \tau^- {\bar \nu_{\tau}}$, other $B{\bar B}$, and $q{\bar q}$ processes are 40, 40, 400, 10 and 5 times larger, respectively, than the full Belle data sample.

\section{Signal Reconstruction}

We first identify events where one of the two $B$ mesons ($B_{\rm tag}$) is reconstructed in one of 1149 exclusive hadronic $B$ decays using a hierarchical multivariate algorithm~\cite{cite-Fulrec} based on the NeuroBayes package~\cite{cite-NeuroBayes}. More than 100 variables are used in this algorithm, including the difference $\Delta E \equiv E_{\rm tag}^* - E_{\rm beam}^*$ between the energy of the reconstructed $B_{\rm tag}$ candidate and the KEKB beam energy in the $e^+ e^-$ center-of-mass system as well as the event shape variables for suppression of $e^+ e^- \rightarrow q{\bar q}$ background. We further require the beam-energy-constrained mass of the $B_{\rm tag}$ candidate $M_{\rm bc} \equiv \sqrt{ E_{\rm beam}^{*2} / c^4 - |\vec{p}_{\rm tag}^{\kern2pt *}|^2 / c^2}$, where $\vec{p}_{\rm tag}^{\kern2pt *}$ denotes the reconstructed $B_{\rm tag}$ three-momentum in the $e^+e^-$ center-of-mass system, to be greater than 5.272~GeV$/c^2$ and $\Delta E$ to lie between $-150$ and $100~{\rm MeV}$. If there are two or more $B_{\rm tag}$ candidates retained after the selection criteria, we select the one with the highest NeuroBayes output value, which is related to the probability that the $B_{\rm tag}$ candidate is correctly reconstructed.

Due to limited knowledge of hadronic $B$ decays, the branching ratios of the $B_{\rm tag}$ decay modes are not perfectly modeled in the MC. It is therefore essential to calibrate the $B_{\rm tag}$ reconstruction efficiency (tagging efficiency) with control data samples. We determine a scale factor for each $B_{\rm tag}$ decay using the method described in Ref.~\cite{cite-Xulnu} based on events where the signal-side $B$ meson candidate ($B_{\rm sig}$) is reconstructed in ${\bar B} \rightarrow D^{(*)} \ell^- {\bar \nu_{\ell}}$ modes. The ratio of measured to expected rates in each decay mode ranges from 0.2 to 1.4, depending on the $B_{\rm tag}$ decay mode, and is 0.68 on average. After the efficiency calibration, the tagging efficiencies are about 0.20\% for charged $B$ mesons and 0.15\% for neutral $B$ mesons.

After $B_{\rm tag}$ selection, we form $B_{\rm sig}$ candidates from the remaining particles not associated with the $B_{\rm tag}$ candidate. Charged particles used to form $B_{\rm sig}$ candidates are reconstructed using information from the SVD and the CDC. The tracks that are not used in $K_S^0$ reconstruction are required to have impact parameters to the interaction point (IP) of less than 0.5~cm (2.0~cm) in the direction perpendicular (parallel) to the $e^+$ beam axis. Charged-particle types are identified by a likelihood ratio based on the response of the sub-detector systems. Identification of $K^{\pm}$ and $\pi^{\pm}$ candidates is done by combining measurements of specific ionization ($dE/dx$) in the CDC, the time of flight from the IP to the TOF counter and the photon yield in the ACC. For $\tau$-daughter $\pi^{\pm}$ candidates, an additional proton veto is required in order to reduce background from the baryonic $B$ decays ${\bar B} \rightarrow D^* {\bar p} n$. The ECL electromagnetic shower shape, track-to-cluster matching at the inner surface of the ECL, the photon yield in the ACC and the ratio of the cluster energy in the ECL to the track momentum measured with the SVD and CDC are used to identify $e^{\pm}$ candidates. Muon candidates are selected based on the comparison of the projected CDC track with interactions in the KLM. To form $K_S^0$ candidates, we combine a pair of oppositely-charged tracks, treated as pions. Three requirements are applied: the reconstructed vertex must be detached from the IP, the momentum vector must point back to the IP, and the invariant mass must be within $\pm$30~MeV/$c^2$ of the nominal $K_S^0$ mass~\cite{cite-PDG}, which corresponds to about 8$\sigma$. (In this section, $\sigma$ denotes the corresponding mass resolution.)

Photons are reconstructed using ECL clusters not matched to charged tracks. Photon energy thresholds of 50, 100 and 150~MeV are used in the barrel, forward-endcap and backward-endcap regions, respectively, of the ECL to reject low-energy background photons, such as those originating from the $e^+ e^-$ beams, and hadronic interactions of particles with material in the detector.

Neutral pions are reconstructed in the decay $\pi^0 \to \gamma\gamma$. For $\pi^0$ candidates from $D$ or $\rho$ decay, we impose the same photon energy thresholds described above. The $\pi^0$ candidate's invariant mass must lie between 115 and 150 MeV/$c^2$, corresponding to about $\pm 3\sigma$ around the nominal $\pi^0$ mass~\cite{cite-PDG}. In order to reduce the number of fake $\pi^0$ candidates, we apply the following $\pi^0$ candidate-selection procedure. The $\pi^0$ candidates are sorted in descending order according to the energy of the most energetic daughter. If a given photon is the most energetic daughter for two or more candidates, they are sorted by the energy of the lower-energy daughter. We then retain the $\pi^0$ candidates whose daughter photons are not shared with a higher-ranked candidate. The remaining $\pi^0$ candidates are used for $D$ or $\rho$ reconstruction described later. For the soft $\pi^0$ from $D^*$ decay, we impose a relaxed photon energy threshold of 22 MeV in all ECL regions, the same requirement for the invariant mass of the two photons, and an energy-asymmetry $A_{\pi^0} = (E_h - E_l)/(E_h + E_l)$ less than 0.6, where $E_h$ and $E_l$ are the energies of the high- and low-energy photon daughters in the laboratory frame. We do not apply the above candidate-selection procedure for the soft $\pi^0$ candidates.

After reconstructing the final-state particles and light mesons, we reconstruct the $D^{(*)}$ candidates using 15 $D$ decay modes: $D^0 \rightarrow K_S^0 \pi^0$, $\pi^- \pi^+$, $K^- \pi^+$, $K^- K^+$, $K^- \pi^+ \pi^0$, $K_S^0 \pi^- \pi^+$, $K_S^0 \pi^- \pi^+ \pi^0$, $K^- \pi^+ \pi^+ \pi^-$, $D^+ \rightarrow K_S^0 \pi^+$, $K_S^0 K^+$, $K_S^0 \pi^+ \pi^0$, $K^- \pi^+ \pi^+$, $K^- K^+ \pi^+$, $K^- \pi^+ \pi^+ \pi^0$, $K_S^0 \pi^+ \pi^+ \pi^-$, and four $D^*$ decay modes: $D^{*0} \rightarrow D^0 \gamma$, $D^0 \pi^0$, $D^{*+} \rightarrow D^+ \pi^0$ and $D^0 \pi^+$. 

The $D$ invariant mass requirements are optimized for each decay mode. For the $D^0$ modes in the $D^{*0}$ candidates, the invariant masses ($m_D$) are required to be within $\pm 2.0\sigma$ ($\pm 1.5\sigma$) of the nominal $D^0$ meson mass~\cite{cite-PDG} for the high (low) signal-to-noise ratio (SNR) modes. For $D^{*+} \rightarrow D^0 \pi^+$ candidates, the $m_D$ requirements are loosened to $\pm 4.0\sigma$ and $\pm 2.0\sigma$ for the high- and low-SNR modes, respectively. The requirements for the $D^+$ candidates are $\pm 2.5\sigma$ for the high-SNR modes and $\pm 1.5\sigma$ for the low-SNR modes around the nominal $D^+$ meson mass~\cite{cite-PDG}. Here, the high-SNR modes are $D^0 \rightarrow K_S^0 \pi^0$, $K^- \pi^+$, $K^- K^+$, $K_S^0 \pi^- \pi^+$, $K^- \pi^+ \pi^+ \pi^-$, $D^+ \rightarrow K_S^0 \pi^+$, $K_S^0 K^+$, $K^- \pi^+ \pi^+$; the low-SNR modes are all remaining $D$ modes. We reconstruct a $D^*$ candidate by combining a $D$ candidate with a $\pi^{\pm}$, $\gamma$ or soft $\pi^0$. The $D^*$ candidates are selected based on the mass difference $\Delta m \equiv m_{D^*} - m_D$, where $m_{D^*}$ denotes the invariant mass of the $D^*$ candidate. The $D^{*0} \rightarrow D^0 \gamma$, $D^{*0} \rightarrow D^0 \pi^0$, $D^{*+} \rightarrow D^+ \pi^0$ and $D^{*+} \rightarrow D^0 \pi^+$ candidates are required to have a $\Delta m$ within $\pm 1.5\sigma$, $\pm 2.0\sigma$, $\pm 2.0\sigma$ and $\pm 3.5\sigma$ of the nominal $\Delta m$.

For the $\tau^- \to \rho^- {\bar \nu_{\tau}}$ candidates, the $\rho$ candidate is formed from the combination of a $\pi^{\pm}$ and a $\pi^0$ with an invariant mass between 0.66 and 0.96~MeV$/c^2$. We then associate a $\pi^{\pm}$ or a $\rho^{\pm}$ candidate (one charged lepton) with the $D^*$ candidate to form signal (normalization) candidates. For the signal mode, square of the momentum transfer,
\begin{eqnarray}
  q^2 &=& (E_{e^+ e^-} - E_{\rm tag} - E_{D^*})^2 / c^2 - (\vec{p}_{e^+ e^-} - \vec{p}_{\rm tag} - \vec{p}_{D^*})^2,\label{eq-q2}
\end{eqnarray}
is required to be greater than 4~GeV$^2/c^2$, where $E$ and $\vec{p}$ denote the energy and the three-momentum specified by the subscript. The subscripts ``$e^+ e^-$'', ``tag'' and ``$D^*$'' stand for the colliding $e^+$ and $e^-$, the $B_{\rm tag}$ candidate and the $D^*$ candidate, respectively. Due to the kinematic constraint in ${\bar B} \rightarrow D^* \tau^- {\bar \nu_{\tau}}$ that $q^2$ is always greater than the square of the $\tau$ mass, almost no signal events exist with $q^2$ below 4~GeV$^2/c^2$. Finally, we require that there be no remaining charged tracks nor $\pi^0$ candidates (except for soft $\pi^0$) in the event.

After the $B_{\rm sig}$ reconstruction procedure is completed, the probability to have multiple candidates (the number of retained candidates) per event is about 9\% (1.09) for the charged $B$ mesons and 3\% (1.03) for the neutral $B$ mesons. Most of the multiple-candidate events are due to the existence of two or more $D^*$ candidates in an event. For the $D^{*0}$ candidates in the charged $B$ meson sample, about 2\% of the events are reconstructed both in the $D^{*0} \rightarrow D^0 \gamma$ and $D^{*0} \rightarrow D^0 \pi^0$ modes. Since the latter mode has a much higher branching fraction, we assign these events to the $D^{*0} \rightarrow D^0 \pi^0$ sample. The contribution of this type of multiple-candidate events is negligibly small in the $D^{*+}$ mode. We then select the most signal-like event as follows. For the $D^{*0} \rightarrow D^0 \gamma$ events, we select the candidate with the most energetic photon associated with the $D^0$. For the $D^{*0} \rightarrow D^0 \pi^0$ and $D^{*+} \rightarrow D^+ \pi^0$ events, we select the candidate with the soft $\pi^0$ having the invariant mass nearest the nominal $\pi^0$ mass. For the $D^{*+} \rightarrow D^0 \pi^+$ events, we select one candidate at random since the multiple-candidate probability is only $\mathcal{O}(0.01\%)$. After the $D^*$ candidate selection, roughly 2\% of the retained events are reconstructed both in the $\tau^- \rightarrow \pi^- \nu_{\tau}$ sample and the $\tau^- \rightarrow \rho^- \nu_{\tau}$ sample. According to the MC study, about 80\% of such signal events are actually $\tau^- \rightarrow \rho^- \nu_{\tau}$ events. We therefore assign these events to the $\tau^- \rightarrow \rho^- \nu_{\tau}$ sample.

\begin{figure}[t!]
  \centering
  \subfigure[]{
    \includegraphics[width=9.5cm]{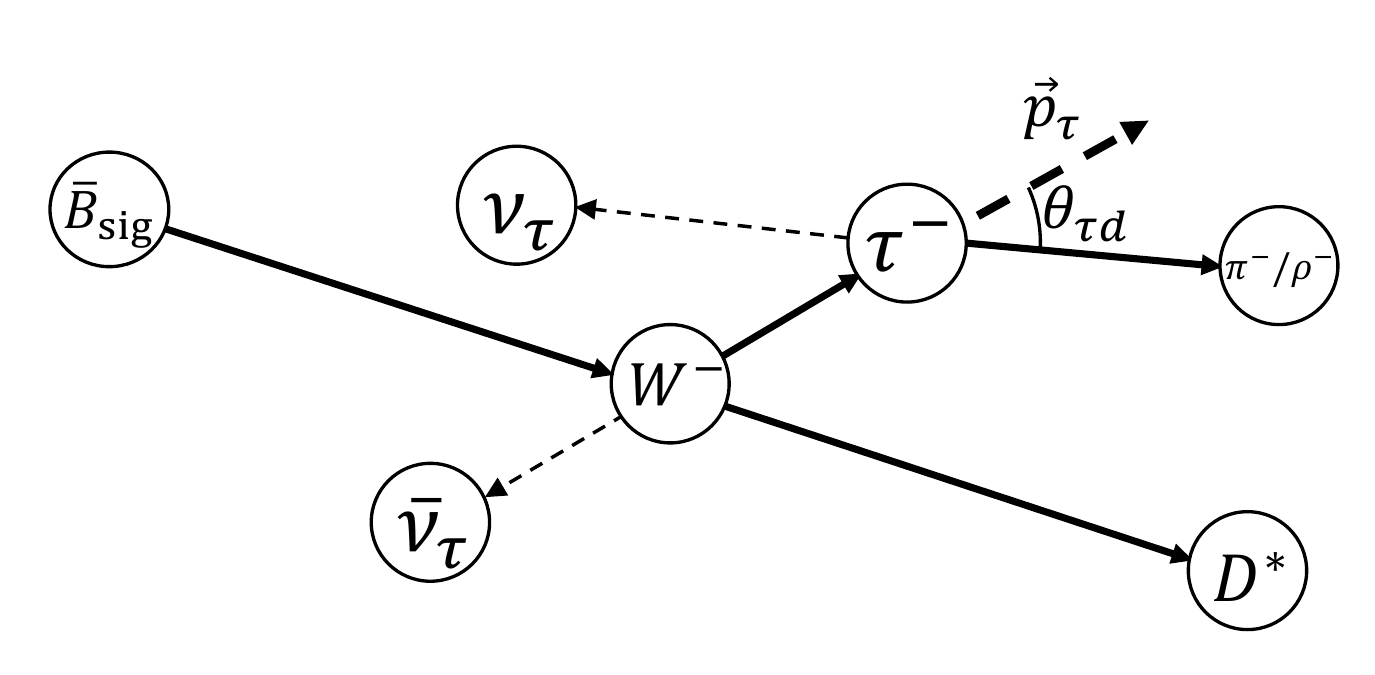}
  }
  \subfigure[]{
    \includegraphics[width=12cm]{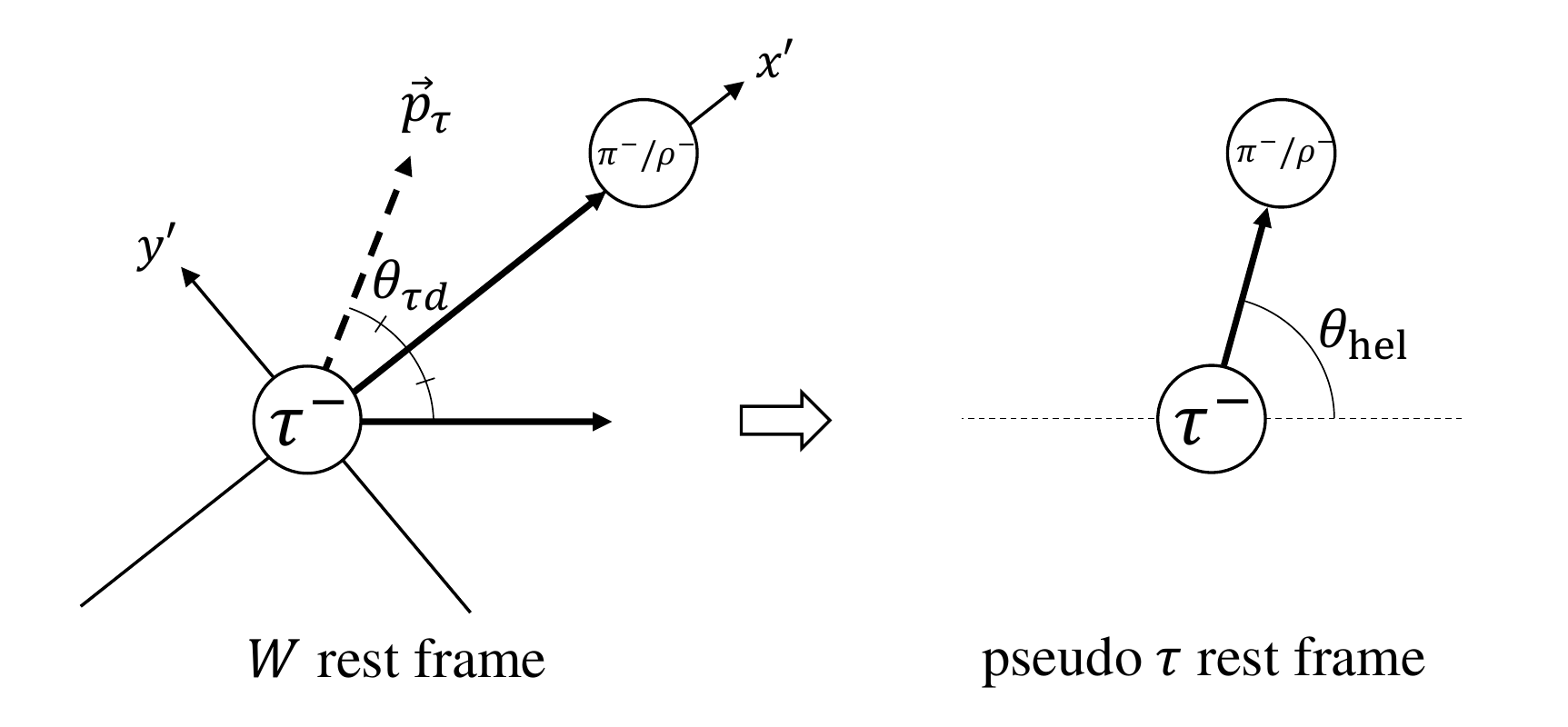}
  }
  \caption{Kinematics of the decay ${\bar B} \rightarrow D^* \tau^- {\bar \nu_{\tau}}$ for $\cos\theta_{\rm hel}$ determination. (a) Decay topology of ${\bar B} \rightarrow D^* \tau^- {\bar \nu_{\tau}}$ in the rest frame of the virtual $W$ boson. The arrows indicate the direction of the momentum vector of each particle. The arrow lengths are not to scale. Neutrino momenta are shown by the thinner-dashed arrows. Every particle except for $\bar{B}_{\rm sig}$ is indicated at the end of the vector. The thicker-dashed arrow expresses the $\tau$ momentum, which is on the cone with the opening angle $\theta_{\tau d}$ around the $\pi^-$ or $\rho^-$ momentum. (b) Transformation from the rest frame of the virtual $W$ boson (left) to the pseudo $\tau$ rest frame (right). The $z'$-axis is perpendicular to the figure plane. The boost axis is taken as the horizontal thicker-solid arrow in the left figure, the direction of which is indicated as the thinner-dashed line in the right figure. As the frame obtained by this boost is not necessarily completely consistent with the $\tau$ rest frame, we name this frame ``pseudo'' $\tau$ rest frame.}
  \label{fig-kinematics}
\end{figure}

\begin{figure}[t!]
  \centering
  \includegraphics[width=10cm]{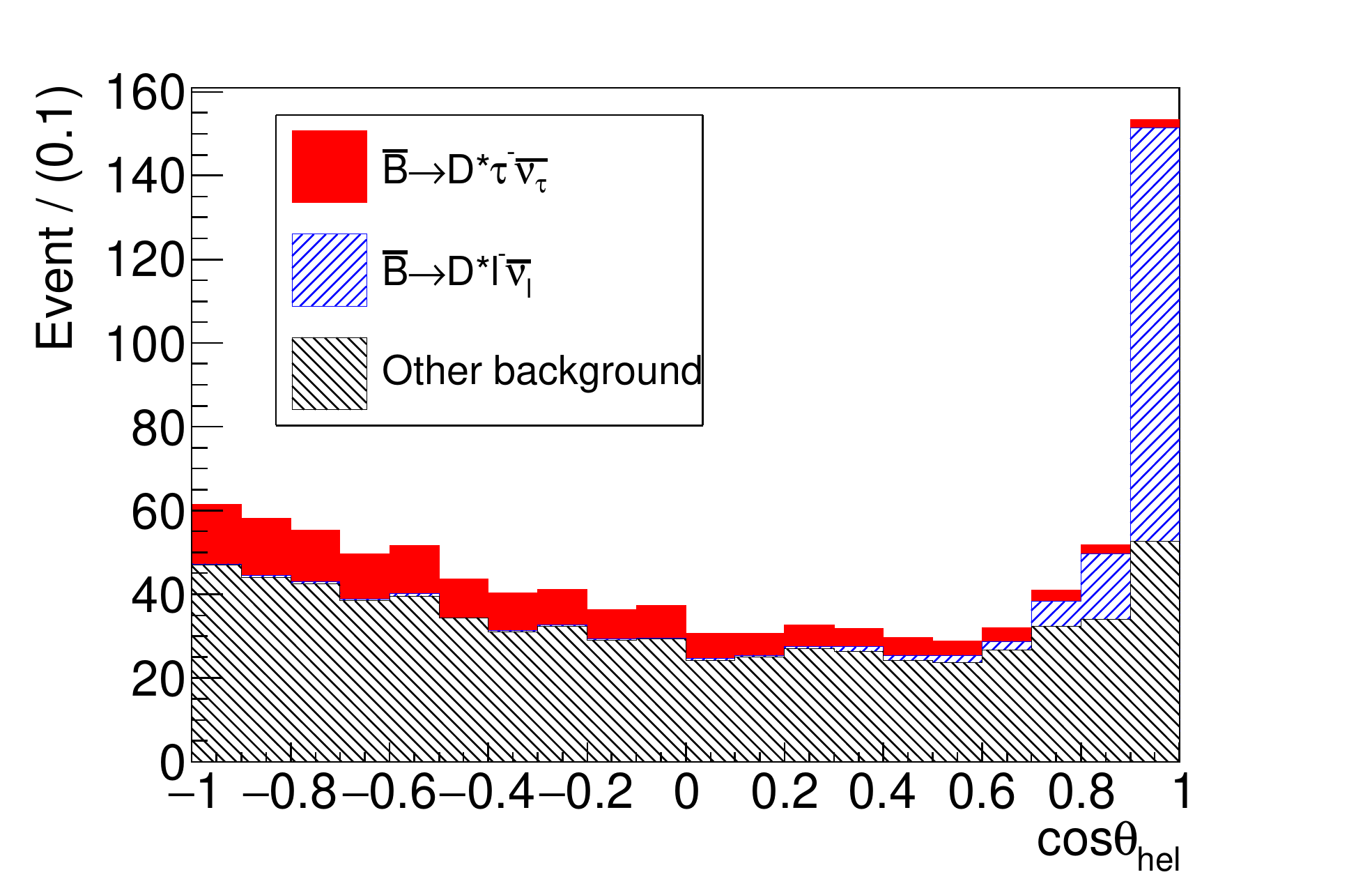}
  \caption{$\cos\theta_{\rm hel}$ distribution for the $B^- \rightarrow D^{*0} \tau^- (\rightarrow \pi^- \nu_{\tau}) {\nu_{\tau}}$ MC sample.}
  \label{fig-coshel}
\end{figure}

In order to measure $P_{\tau}$, the $\cos\theta_{\rm hel}$ distribution must be reconstructed. This is challenging, as the $\tau$ momentum vector is not fully determined. Instead of $\cos\theta_{\rm hel}$, we measure the cosine of the angle $\theta_{\tau d}$ between the momenta of the $\tau$ lepton and its daughter meson in the rest frame of the virtual $W$ boson,
\begin{eqnarray}
  \cos\theta_{\tau d} &=& \frac{2 E_{\tau} E_d - m_{\tau}^2 c^4 - m_d^2 c^4}{2 |\vec{p}_{\tau}||\vec{p}_d| c^2},
\end{eqnarray}
as shown in Fig.~\ref{fig-kinematics} (a). Here, $E$ and $\vec{p}$ denote the energy and the three-momentum of a particle specified by the subscript, where $\tau$ and $d$ represent the $\tau$ lepton and its daughter meson, respectively. This angle is equivalent to $\theta_{\rm hel}$ in this frame. The rest frame of the virtual $W$ is obtained from its three-momentum
\begin{eqnarray}
  \vec{p}_W &=& \vec{p}_{e^+ e^-} - \vec{p}_{\rm tag} - \vec{p}_{D^*} = 0.
\end{eqnarray}
In this frame of reference, the magnitude of the $\tau$ momentum is determined only by $q^2$ since the $\tau$ lepton is emitted in the two-body decay of the static virtual $W$ boson. Therefore $|\vec{p}_{\tau}|$ is calculated as
\begin{eqnarray}
  |\vec{p}_{\tau}| &=& \frac{q^2 - m_{\tau}^2/c^2}{2 \sqrt{q^2}}.
\end{eqnarray}
We only accept events, for which $|\cos\theta_{\tau d}| < 1$. Here, more than 97\% of the reconstructed signal events are retained.

Due to limited kinematic constraints, one degree of freedom of the $\tau$ momentum direction is not determined. However, the cone around $\vec{p}_d$ with an angle of $\theta_{\tau d}$, on which $\vec{p}_{\tau}$ lies, is rotationally symmetric and therefore all directions on this cone are equivalent. With this in mind, as shown in Fig.~\ref{fig-kinematics} (b), we take the new right-handed $x'y'z'$ coordinate such that the $x'$-axis corresponds to the direction of the $\vec{p}_d$, and set $\vec{p}_{\tau} = (|\vec{p}_{\tau}| \cos\theta_{\tau d}, |\vec{p}_{\tau}| \sin\theta_{\tau d}, 0)$. The system is boosted to the pseudo $\tau$ rest frame with $\vec{p}_{\tau} = 0$, where the correct value of $\cos\theta_{\rm hel}$ is obtained.

As shown in Fig.~\ref{fig-coshel}, there are many ${\bar B} \rightarrow D^* \ell^- {\bar \nu_{\ell}}$ background events that peak around $\cos\theta_{\rm hel} = 1$, which corresponds to $M_{\rm miss}^2 \sim 0$, in the $\tau^- \rightarrow \pi^- \nu_{\tau}$ sample. This peak arises from low-momentum muons that do not reach the KLM and are therefore misreconstructed as pions. To mitigate this background, we only use the region $\cos\theta_{\rm hel} < 0.8$ in the fit, in which 94\% (81\%) of signal events are contained with the SM $P_{\tau}$ of $-0.497$ (maximum $P_{\tau}$ of $+1$).

\section{Background Separation and Calibration}\label{sec-background}

In order to separate signal and normalization events from background, we use the variable $E_{\rm ECL}$, the summed energy of ECL clusters not used in the reconstruction of the $B_{\rm sig}$ and $B_{\rm tag}$ candidates. This is a useful variable for the signal extraction since the $E_{\rm ECL}$ shape is less affected by changes in kinematics due to NP. The variable $M_{\rm miss}^2$ is additionally used for normalization events, and is defined as
\begin{eqnarray}
  M_{\rm miss}^2 &=& (E_{e^+ e^-} - E_{\rm tag} - E_{D^*} - E_{\ell})^2 / c^4 - (\vec{p}_{e^+ e^-} - \vec{p}_{\rm tag} - \vec{p}_{D^*} - \vec{p}_{\ell})^2 / c^2,
\end{eqnarray} 
where $E_{\ell}$ and $\vec{p}_{\ell}$ are the energy and the three-momentum, respectively, of the charged lepton; the other variables in this formula are defined in Eq.~\ref{eq-q2}. Due to its narrow concentration near $M_{\rm miss}^2 = 0$, this variable is ideal for measuring normalization events. In the fit, we use the distributions obtained from MC for the PDFs. The $E_{\rm ECL}$ shape for the signal component is validated using the normalization sample, which is more than 20 times larger than the signal sample. In this comparison, we find good agreement between the data and the MC distributions. In the $M_{\rm miss}^2$ comparison for the normalization sample, the $M_{\rm miss}^2$ resolution in the data sample is slightly worse than in the MC sample. We therefore broaden the $M_{\rm miss}^2$ peak width of the PDFs to match that of the data sample.

The most significant background contribution is from events with incorrectly-reconstructed $D^*$ (denoted ``fake $D^*$ events''). Since the combinatorial fake $D^*$ background processes are difficult to be modeled precisely in the MC, we compare the PDF shapes of these events in $\Delta m$ sideband regions. The sideband regions 50--500~MeV$/c^2$, 135--190~MeV$/c^2$, 135--190~MeV$/c^2$ and 140--500~MeV$/c^2$ are chosen for $D^{*0} \rightarrow D^0 \gamma$, $D^{*0} \rightarrow D^0 \pi^0$, $D^{*+} \rightarrow D^+ \pi^0$ and $D^{*+} \rightarrow D^0 \pi^+$, respectively, while excluding about $\pm 3.5\sigma$ around the nominal $\Delta m$. These sideband regions contain 5--50 times more events than the signal region. While we find good agreement of the $E_{\rm ECL}$ shapes between the data and the MC for the signal sample, we observe a slight discrepancy in the $M_{\rm miss}^2$ distributions of the $D^{*0} \rightarrow D^0 \gamma$ and $D^{*0} \rightarrow D^0 \pi^0$ modes for the normalization sample. The $M_{\rm miss}^2$ discrepancy is therefore corrected based on this comparison. In both samples, since up to 20\% of the yield discrepancies are observed, the fake $D^*$ yields are scaled by the yield ratios of the data to the MC in the $\Delta m$ sideband regions.

Semileptonic decays to excited charm modes, ${\bar B} \rightarrow D^{**} \ell^- {\bar \nu_{\ell}}$ and ${\bar B} \rightarrow D^{**} \tau^- {\bar \nu_{\tau}}$, comprise an important background category as they have a similar decay topology to the signal events. In addition, background events from various types of hadronic $B$ decays wherein some particles are not reconstructed are significant in this analysis since there are only hadrons and two neutrinos in the final state of the signal mode. Because there are many unmeasured exclusive decay modes of ${\bar B} \rightarrow D^{**} \ell^- {\bar \nu_{\ell}}$, ${\bar B} \rightarrow D^{**} \tau^- {\bar \nu_{\tau}}$ and hadronic $B$ decays, we determine their yields in the fit. With one exception, we sum all the exclusive decays of these background categories into common yield parameters. The exception is the decay to two $D$ mesons, such as ${\bar B} \rightarrow D^* D_s^{*-}$ and ${\bar B} \rightarrow D^* {\bar D^*} K^-$, since these are experimentally well measured: we fix their yields based on the world-average branching fractions~\cite{cite-PDG}.

In addition to the yield determination, the PDF shape of these background must be taken into account, as a change in the $B$ decay composition may modify the $E_{\rm ECL}$ shape and thereby introduce biases in the measurement of $R(D^*)$ and $P_{\tau}$. If a background $B$ decay contains a $K_L^0$ in the final state, it may peak in the $E_{\rm ECL}$ signal region. We correct the branching fractions of the ${\bar B} \rightarrow D^* \pi^- K_L^0$ and ${\bar B} \rightarrow D^* K^- K_L^0$ modes in the MC using the measured values~\cite{cite-PDG,cite-DstKKL}. We do not apply branching fraction corrections for the other decays with $K_L^0$ since they are relatively minor. However, we change the relative yield from 0\% to 200\% to estimate systematic uncertainties, as discussed in Sec.~\ref{sec-syst}.

\begin{table}[t!]
  \centering
  \caption{List of the calibration factors for each calibration sample, which are used to correct the amount of each hadronic $B$ background in the MC. These calibration factors are obtained from the yield comparison between the data and the MC with the $\Delta E^{\rm sig}$ or $M_{\rm bc}^{\rm sig}$ distributions. The errors on the calibration factors arise from statistics of the calibration samples.}
  \vspace{3mm}
  \begin{tabular}{@{\hspace{0.5cm}}l@{\hspace{0.5cm}}|@{\hspace{0.5cm}}c@{\hspace{0.5cm}}@{\hspace{0.5cm}}c@{\hspace{0.5cm}}}
    \hline
    \hline
    $B$ decay mode & $B^-$ & ${\bar B^0}$\\
    \hline
    $D^* \pi^- \pi^- \pi^+$             & $< 0.49$                 & $0.57 ^{+0.67} _{-0.49}$\\ 
    $D^* \pi^- \pi^- \pi^+ \pi^0$       & $0.31 ^{+0.43} _{-0.40}$ & $0.59 ^{+0.45} _{-0.39}$\\ 
    $D^* \pi^- \pi^- \pi^+ \pi^0 \pi^0$ & $2.68 ^{+2.74} _{-2.55}$ & $2.44 ^{+5.88} _{-2.24}$\\ 
    $D^* \pi^- \pi^0$                   & $0.06 ^{+0.33} _{-0.28}$ & $< 0.46$\\ 
    $D^* \pi^- \pi^0 \pi^0$             & $0.09^{+1.04} _{-0.98}$  & $1.63 ^{+0.74} _{-0.69}$\\ 
    $D^* \pi^- \eta$                    & $0.24 ^{+0.21} _{-0.19}$ & $0.15 ^{+0.16} _{-0.10}$\\
    $D^* \pi^- \eta \pi^0$              & $0.74 ^{+0.79} _{-0.75}$ & $0.91 ^{+1.11} _{-0.93}$\\ 
    \hline
    \hline
  \end{tabular}
  \label{tab-calib}
\end{table}

Other types of hadronic $B$ decay background often contain neutral particles such as $\pi^0$ or $\eta$ or pairs of charged particles. We calibrate the amount of hadronic $B$ decays in the MC based on control data samples by reconstructing seven final states with the signal-side particles: ${\bar B} \rightarrow D^* \pi^- \pi^- \pi^+$, ${\bar B} \rightarrow D^* \pi^- \pi^- \pi^+ \pi^0$, ${\bar B} \rightarrow D^* \pi^- \pi^- \pi^+ \pi^0 \pi^0$, ${\bar B} \rightarrow D^* \pi^- \pi^0$, ${\bar B} \rightarrow D^* \pi^- \pi^0 \pi^0$, ${\bar B} \rightarrow D^* \pi^- \eta$ and ${\bar B} \rightarrow D^* \pi^- \eta \pi^0$. Candidate $\eta$ mesons are reconstructed using pairs of photons with an invariant mass ranging from 500 to 600~MeV$/c^2$. We then take the yield ratios between the data and the MC for $q^2 > 4~{\rm GeV}^2/c^2$ and $|\cos\theta_{\rm hel}| < 1$, which is the same requirement as in the signal sample, with the signal-side energy difference $\Delta E^{\rm sig}$ or the beam-energy-constrained mass $M_{\rm bc}^{\rm sig}$ of the $B_{\rm sig}$ candidate. These ratios are used as yield calibration factors. If there is no observed event in the calibration sample, we assign a 68\% confidence level upper limit on the yield. The obtained factors are summarized in Table~\ref{tab-calib}. Additionally, we correct the branching fractions of the decays $B^- \rightarrow D^{*+} \pi^- \pi^- \pi^0$, ${\bar B} \rightarrow D^* \omega \pi^-$ and ${\bar B} \rightarrow D^* {\bar p} n$ based on Refs.~\cite{cite-PDG,cite-Dstomegapi}.

About 80\% of the hadronic $B$ background is covered by the calibrations discussed above. We discuss the systematic uncertainties on our observables due to the uncertainties of the calibration factors in Sec.~\ref{sec-syst}.

\section{Maximum Likelihood Fit}

We perform an extended binned maximum likelihood fit to the $E_{\rm ECL}$ and the $M_{\rm miss}^2$ distributions for the signal- and the normalization-candidate samples, respectively. In order to extract $P_{\tau}$, we divide the signal sample into two $\cos\theta_{\rm hel}$ regions: $\cos\theta_{\rm hel} > 0$ (forward) and $\cos\theta_{\rm hel} < 0$ (backward). According to Eq.~\ref{eq-coshel}, the asymmetry of the number of signal events between the forward and the backward regions is proportional to $P_{\tau}$.

In the fit, we divide the ${\bar B} \rightarrow D^* \tau^- {\bar \nu_{\tau}}$ component into three groups.

\begin{description}
\item[Signal]\mbox{}\\
  Correctly-reconstructed signal events, which originate from $\tau^- \rightarrow \pi^- (\rho^-) \nu_{\tau}$ events reconstructed correctly as the $\tau^- \rightarrow \pi^- (\rho^-) \nu_{\tau}$ sample, are categorized in this component, and are used for the determination of $R(D^*)$ and $P_{\tau}$.
\item[$\rho \leftrightarrow \pi$ cross feed]\mbox{}\\
  Cross-feed events where the decay $\tau^- \rightarrow \rho^- \nu_{\tau}$ is reconstructed in the $\tau^- \rightarrow \pi^- \nu_{\tau}$ mode due to the misreconstruction of one $\pi^0$, or events where the decay $\tau^- \rightarrow \pi^- \nu_{\tau}$ is reconstructed in the  $\tau^- \rightarrow \rho^- \nu_{\tau}$ mode by adding a random $\pi^0$, comprise this component. As these events originate from ${\bar B} \rightarrow D^* \tau^- {\bar \nu_{\tau}}$, they can be used in the $R(D^*)$ determination. They also have some sensitivity to $P_{\tau}$; however, $\cos\theta_{\rm hel}$ is distorted. The measured $P_{\tau}$ from the distorted $\cos\theta_{\rm hel}$ distribution is mapped to the correct value of $P_{\tau}$ using MC information.
\item[Other $\tau$ cross feed]\mbox{}\\
Events from other $\tau$ decays also can contribute to the signal sample. They originate mainly from $\tau^- \rightarrow \pi^- \pi^0 \pi^0 \nu_{\tau}$ with two missing $\pi^0$ mesons and $\tau^- \rightarrow \mu^- {\bar \nu_{\mu}} \nu_{\tau}$ with a low-momentum muon. The fraction of these two cross-feed components are, respectively, 11\% and 73\% in the $\tau^- \rightarrow \pi^- \nu_{\tau}$ mode and 69\% and 14\% in the $\tau^- \rightarrow \rho^- \nu_{\tau}$ mode. These modes are less sensitive to $P_{\tau}$ since the heavy $a_1$ mass makes the $\alpha$ in Eq.~(\ref{eq-alpha}) almost equal to 0, while events with two neutrinos in the $\tau^- \rightarrow \mu^- {\bar \nu_{\mu}} \nu_{\tau}$ mode wash out the $P_{\tau}$ information.
\end{description}
The relative contribution from the three ${\bar B} \rightarrow D^* \tau^- {\bar \nu_{\tau}}$ components are fixed using the MC simulation sample, which contains 40 times more events than the full Belle data sample.

The parameterization of $R(D^*)$ is
\begin{eqnarray}
  R(D^*) &=& \frac{1}{\mathcal{B}_{\tau}}\frac{\epsilon_{\rm norm}}{\epsilon_{\rm sig}}\frac{N_{\rm sig, F} + N_{\rm sig, B}}{N_{\rm norm}},
\end{eqnarray}
where $\mathcal{B}_{\tau}$ denotes the branching fraction of $\tau^- \rightarrow \pi^- \nu_{\tau}$ or $\tau^- \rightarrow \rho^- \nu_{\tau}$, and $\epsilon_{\rm sig}$ and $\epsilon_{\rm norm}$ are the efficiencies for the signal and the normalization mode, respectively. The observed yields are expressed by $N_{\rm sig, F(B)}$ and $N_{\rm norm}$ for the signal in the forward (backward) region and the normalization, respectively. The polarization is represented by
\begin{eqnarray}
  P_{\tau} &=& \frac{2}{\alpha}\frac{N_{\rm sig, F} - N_{\rm sig, B}}{N_{\rm sig, F} + N_{\rm sig, B}}.
\end{eqnarray}
Due to detector effects, the extracted value deviates from the true $P_{\tau}$. This detector bias is taken into account with a linear function that relates the true $P_{\tau}$ to the extracted $P_{\tau}$. The linear function, which is called the $P_{\tau}$ correction function in this paper, is determined using several MC sets of type-II 2HDM~\cite{cite-2HDM-II}. In this model, $P_{\tau}$ varies between $-0.6$ and $+1.0$ as a function of the theoretical parameter $\tan\beta / m_{H^+}$, where $\tan\beta$ denotes the ratio of the vacuum expectation values of the two Higgs doublets coupling to up-type and down-type quarks and $m_{H^+}$ is the mass of the charged Higgs boson. We then extrapolate the obtained $P_{\tau}$ correction function to $P_{\tau} = -1$.

For the background, we have four components.
\begin{description}
\item[${\bar B} \rightarrow D^* \ell^- {\bar \nu_{\ell}}$]\mbox{}\\
  The decay ${\bar B} \rightarrow D^* \ell^- {\bar \nu_{\ell}}$ contaminates the signal sample due to the misassignment of the lepton as a pion. We fix the ${\bar B} \rightarrow D^* \ell^- {\bar \nu_{\ell}}$ yield in the signal sample from the fit to the $M_{\rm miss}^2$ distribution in the normalization sample.  
\item[${\bar B} \rightarrow D^{**} \ell^- {\bar \nu_{\ell}}$ and hadronic $B$ decays]\mbox{}\\
  As discussed in the previous section, we float the sum of the yields of ${\bar B} \rightarrow D^{**} \ell^- {\bar \nu_{\ell}}$ and hadronic $B$ decays except for the well-determined two-body $D$ final states in the fit. The yield parameters are independent for each sample: $(B^-, {\bar B^0}) \otimes (\pi^- \nu_{\tau}, \rho^- \nu_{\tau}) \otimes (forward, backward)$.
\item[Continuum]\mbox{}\\
  Continuum events from $e^+ e^- \rightarrow q{\bar q}$ process provide a minor contribution. As the size of the contribution is only $\mathcal{O}(0.1\%)$, we fix the yield using the MC expectation.
\item[Fake $D^*$]\mbox{}\\
  All events containing fake $D^*$ candidates are categorized in this component. The yield is fixed from a comparison of the data and the MC in the $\Delta m$ sideband regions.
\end{description}

\section{Systematic Uncertainties}\label{sec-syst}

\begin{table}[t!]
  \centering
  \caption{The systematic uncertainties in $R(D^*)$ and $P_{\tau}$, where the values for $R(D^*)$ are relative errors. The group ``common sources'' identifies the common systematic uncertainty sources in the signal and the normalization modes, which cancel to a good extent in the ratio of these samples. The reason for the incomplete cancellation is described in the text.}
  \vspace{3mm}
  \begin{tabular}{@{\hspace{0.5cm}}l@{\hspace{0.5cm}}|@{\hspace{0.5cm}}c@{\hspace{0.5cm}}@{\hspace{0.5cm}}c@{\hspace{0.5cm}}}
    \hline
    \hline
    Source & $R(D^*)$ & $P_{\tau}$\\
    \hline
    Hadronic $B$ composition                 & $^{+7.8\%}_{-6.9\%}$ & $^{+0.14}_{-0.11}$\\
    MC statistics for each PDF shape         & $^{+3.5\%}_{-2.8\%}$ & $^{+0.13}_{-0.11}$\\
    Fake $D^*$ PDF shape                     & 3.0\% & 0.010\\
    Fake $D^*$ yield                         & 1.7\% & 0.016\\
    ${\bar B} \rightarrow D^{**} \ell^- {\bar \nu_{\ell}}$ & 2.1\% & 0.051\\
    ${\bar B} \rightarrow D^{**} \tau^- {\bar \nu_{\tau}}$ & 1.1\% & 0.003\\
    ${\bar B} \rightarrow D^* \ell^- {\bar \nu_{\ell}}$ & 2.4\% & 0.008\\
    $\tau$ daughter and $\ell^-$ efficiency  & 2.1\% & 0.018\\
    MC statistics for efficiency calculation & 1.0\% & 0.018\\
    EvtGen decay model                       & $^{+0.8\%}_{-0.0\%}$ & $^{+0.016}_{-0.000}$\\
    Fit bias                                 & 0.3\% & 0.008\\
    $\mathcal{B}(\tau^- \rightarrow \pi^- \nu_{\tau})$ and $\mathcal{B}(\tau^- \rightarrow \rho^- \nu_{\tau})$ & 0.3\% & 0.002\\
    $P_{\tau}$ correction function           & 0.1\% & 0.018\\
    \hline
    \multicolumn{3}{c}{Common sources}\\
    \hline
    Tagging efficiency correction            & 1.4\% & 0.014\\
    $D^*$ reconstruction                     & 1.3\% & 0.007\\
    $D$ sub-decay branching fractions        & 0.7\% & 0.005\\
    Number of $B{\bar B}$                    & 0.4\% & 0.005\\
    \hline
    Total systematic uncertainty & $^{+10.4\%}_{-9.5\%}$ & $^{+0.20}_{-0.17}$\\
    \hline
    \hline
  \end{tabular}
  \label{tab-syst}
\end{table}

We estimate systematic uncertainties by varying each possible uncertainty source such as the PDF shape and the signal reconstruction efficiency with the assumption of a Gaussian error, unless otherwise stated. In several trials, we change each parameter at random, repeat the fit, and then take the mean shifts of $R(D^*)$ and $P_{\tau}$ from all such trials as the corresponding systematic uncertainty that is enumerated in Table~\ref{tab-syst}.

The most significant systematic uncertainty, arising from the hadronic $B$ decay composition, is estimated as follows. Uncertainties of each $B$ decay fraction in the hadronic $B$ decay background are taken from the experimentally-measured branching fractions or estimated from the uncertainties in the calibration factors discussed in Sec.~\ref{sec-background}. For components with no experimentally-measured branching fractions and not covered by the control samples, we vary their contribution continuously from $0\%$ to $200\%$ of the MC expectation and take the maximum shifts of $R(D^*)$ and $P_{\tau}$ as the systematic uncertainties.

The limited MC sample size used in the construction of the PDFs is also a major systematic uncertainty source. We estimate this by regenerating the PDFs for each component and each sample using a toy MC approach based on the original PDF shapes. The same number of events are generated to account for the statistical fluctuation.

The PDF shape of the fake $D^*$ component has been validated by comparing the data and the MC in the $\Delta m$ sideband region. However, a slight fluctuation from the decay ${\bar B} \rightarrow D \tau^- {\bar \nu_{\tau}}$, which is a peaking background in the fake $D^*$ component, may have a significant impact on the signal yield as this component has almost the same shape as the signal mode. To be conservative, we incorporate an additional uncertainty by varying the contribution of the ${\bar B} \rightarrow D \tau^- {\bar \nu_{\tau}}$ component within the current uncertainties of the experimental averages~\cite{cite-PDG}: $\pm 32\%$ for $B^- \rightarrow D \tau^- {\bar \nu_{\tau}}$ and $\pm 21\%$ for ${\bar B^0} \rightarrow D \tau^- {\bar \nu_{\tau}}$. We take the theoretical uncertainty on the $\tau$ polarization of the ${\bar B} \rightarrow D \tau^- {\bar \nu_{\tau}}$ mode into account, which is found to be 0.002 for $P_{\tau}$ and negligibly small. In addition, we estimate a systematic uncertainty due to the small $M_{\rm miss}^2$ shape correction for the fake $D^*$ component of the normalization sample: this is 0.15\% (0.001) for $R(D^*)$ ($P_{\tau}$).

The fake $D^*$ yield, fixed using the $\Delta m$ sideband, has an uncertainty that arises from the statistical uncertainties of the yield scale factors.

The uncertainty of the decays ${\bar B} \rightarrow D^{**} \ell^- {\bar \nu_{\ell}}$ are twofold: the indeterminate composition of each $D^{**}$ state and the uncertainty in the form-factor parameters used for the MC sample production. The composition uncertainty is estimated based on uncertainties of the branching fractions: $\pm 6\%$ for ${\bar B} \rightarrow D_1 (\rightarrow D^* \pi) \ell {\bar \nu_{\ell}}$, $\pm 12\%$ for ${\bar B} \rightarrow D_2^* (\rightarrow D^* \pi) \ell {\bar \nu_{\ell}}$, $\pm 24\%$ for ${\bar B} \rightarrow D'_1 (\rightarrow D^* \pi \pi) \ell {\bar \nu_{\ell}}$ and $\pm 17\%$ for ${\bar B} \rightarrow D_0^* (\rightarrow D^* \pi) \ell {\bar \nu_{\ell}}$. If the experimentally-measured branching fractions are not applicable, we vary the branching fractions continuously from $0\%$ to $200\%$ in the MC expectation. We estimate an uncertainty arising from the LLSW model parameters by changing the correction factors within the parameter uncertainties and obtain 0.5\% and 0.016 for $R(D^*)$ and $P_{\tau}$, respectively.

The uncertainty due to limited knowledge of the decays ${\bar B} \rightarrow D^{**} \tau^- {\bar \nu_{\tau}}$ is estimated separately by varying the branching fractions. Since there are no experimental measurements of these decays, the branching fractions are varied continuously from 0\% to 200\% in the MC expectation. As $\mathcal{B}({\bar B} \rightarrow D^{**} \tau^- {\bar \nu_{\tau}})$ is constrained by the branching fraction of the inclusive semitauonic decay $\mathcal{B}({\bar B} \rightarrow X_c \tau^- {\bar \nu_{\tau}}) = (2.41 \pm 0.23)\%$~\cite{cite-PDG}, which is smaller than the sum of the branching fractions $\mathcal{B}({\bar B} \rightarrow D \tau^- {\bar \nu_{\tau}})$ and $\mathcal{B}({\bar B} \rightarrow D^* \tau^- {\bar \nu_{\tau}})$, we conclude that our assumption is sufficiently conservative.

The uncertainties due to the HQET form-factor parameters in the normalization mode ${\bar B} \rightarrow D^* \ell^- {\bar \nu_{\ell}}$ are estimated using the uncertainties in the world-average values~\cite{cite-HFAG}. In addition, the uncertainty arising from the small $M_{\rm miss}^2$ shape correction for the normalization sample is estimated as an uncertainty related to ${\bar B} \rightarrow D^* \ell^- {\bar \nu_{\ell}}$: 0.4\% (0.008) for $R(D^*)$ ($P_{\tau}$).

The uncertainties on the reconstruction efficiencies of the $\tau$-daughter particles in the signal sample and the charged leptons in the normalization sample are also considered. Here, the uncertainties on the particle identification efficiencies for $\pi^{\pm}$ and $\ell^{\pm}$ and the reconstruction efficiency for $\pi^0$ are measured with control samples: the $D^{*+} \rightarrow D^0(\rightarrow K^- \pi^+)\pi^+$ sample for $\pi^{\pm}$, the $\tau^- \rightarrow \pi^- \pi^0 \nu_{\tau}$ sample for $\pi^0$ and the $\gamma \gamma \rightarrow \ell^+ \ell^-$ for charged leptons. The sample $J/\psi \rightarrow \ell^+ \ell^-$ from $B$ decays is also used in order to account for the difference in multiplicity between two-photon events and $B$ decay events. Estimated uncertainties are as follows: 0.5\% and 0.003 for $\pi^{\pm}$, 0.6\% and 0.004 for $\pi^0$, 1.6\% and 0.004 for charged leptons (the first and the second values corresponding to uncertainties of $R(D^*)$ and $P_{\tau}$, respectively).

Reconstruction efficiencies of the three ${\bar B} \rightarrow D^* \tau^- {\bar \nu_{\tau}}$ components are estimated using MC. The efficiencies have uncertainties arising from the statistics of the signal MC and are varied independently for each component. Here, the uncertainty of the fraction of the three ${\bar B} \rightarrow D^* \tau^- {\bar \nu_{\tau}}$ components are also taken into account.

The ${\bar B} \rightarrow D^* \tau^- {\bar \nu_{\tau}}$ decay generator of EvtGen, based on the HQET form factors and implemented by Belle, neglects the interference between the amplitudes of left- and right-handed $\tau$ leptons. This mis-models the decay topology and so affects the signal reconstruction efficiency. We compare the extracted $R(D^*)$ and $P_{\tau}$ between this model and an alternate model based on the ISGW form factors, and take differences of $R(D^*)$ and $P_{\tau}$ as systematic uncertainties.

Other small uncertainties arise due to fit bias arising from the $E_{\rm ECL}$ bin width selection, which is estimated by comparing the extracted values of $R(D^*)$ and $P_{\tau}$ in the nominal fit (with a bin width of 0.05~GeV) with the values obtained in a wide-bin fit (bin width of 0.1~GeV); the branching fractions of the $\tau$ lepton decays; and errors on the parameters of the $P_{\tau}$ correction function.

\begin{figure}[t!]
  \centering
  \includegraphics[width=15cm]{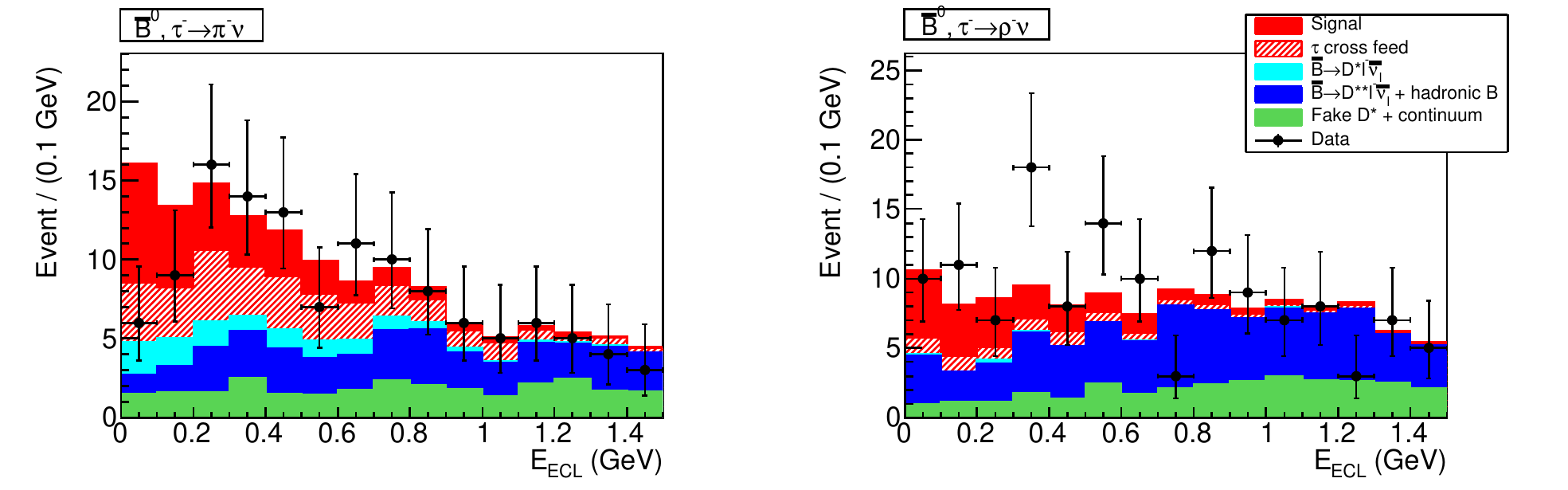}
  \includegraphics[width=15cm]{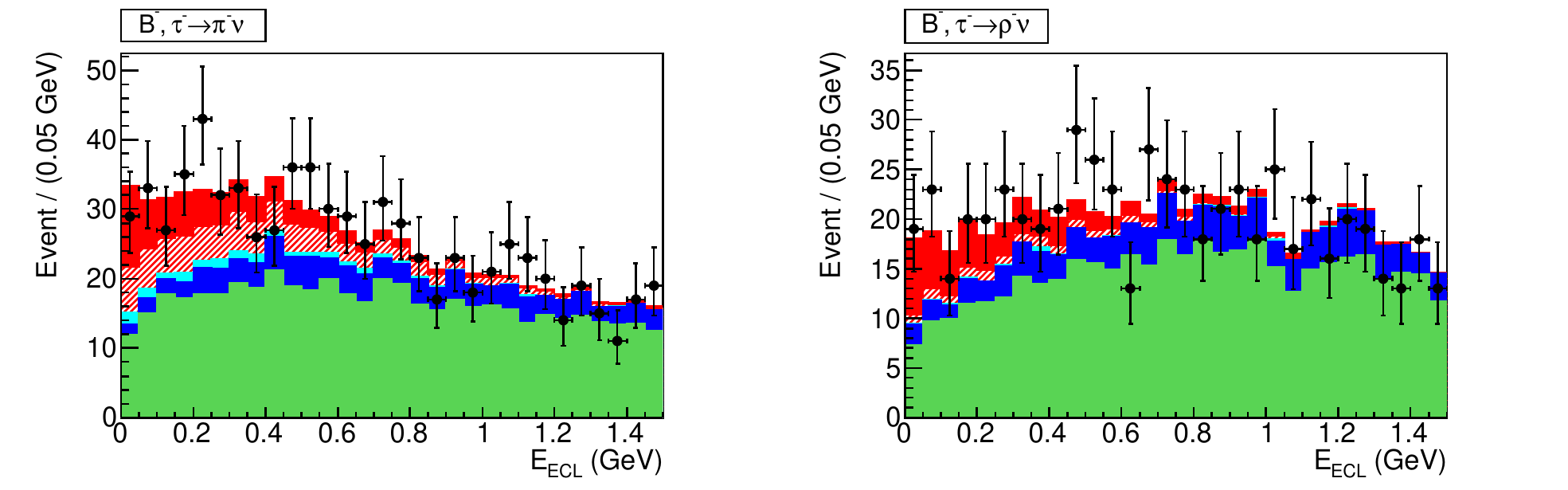}
  \includegraphics[width=15cm]{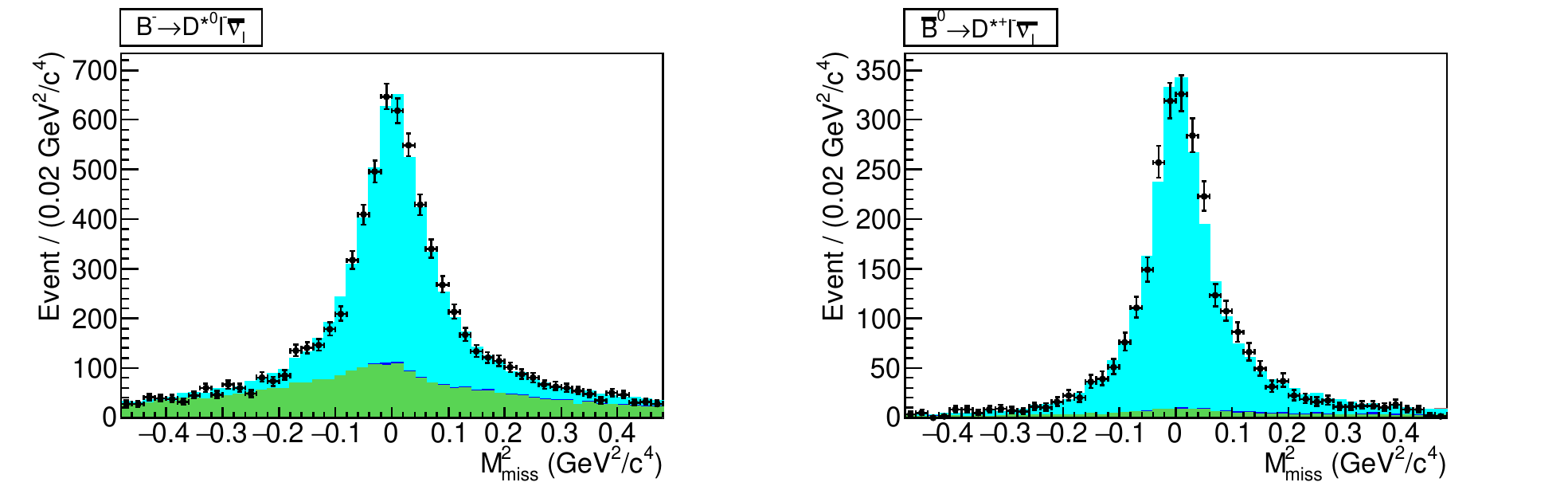}
  \caption{Fit results to the four signal and two normalization samples. For the signal sample, the projection onto the $E_{\rm ECL}$ axis is illustrated. The red-hatched ``$\tau$ cross feed'' combines the $\rho \leftrightarrow \pi$ cross-feed and the other $\tau$ cross-feed components.}
  \label{fig-fitresult}
\end{figure}

In addition, common uncertainty sources between the signal sample and the normalization sample are also estimated in this analysis, although they largely cancel at first order in the branching fraction ratio. This is due to the fact that the background yields are partially fixed from the MC expectation. Here, uncertainties on the number of $B{\bar B}$ (1.9\%) events, tagging efficiencies (4.7\%), branching fractions of the $D$ decays (3.4\%) and $D^*$ reconstruction efficiency (4.8\%) are evaluated for their impact on the final measurements. For the $D^*$ reconstruction efficiency, the uncertainty originates from reconstruction efficiencies of $K_S^0$, $\pi^0$, $K^{\pm}$ and $\pi^{\pm}$, and is therefore correlated with the efficiency uncertainty of the $\tau$-daughter particles containing $\pi^{\pm}$ and $\pi^0$. This correlation is taken into account in the total systematic uncertainties shown in Table~\ref{tab-syst}.

\begin{figure}[t!]
  \centering
  \includegraphics[width=15cm]{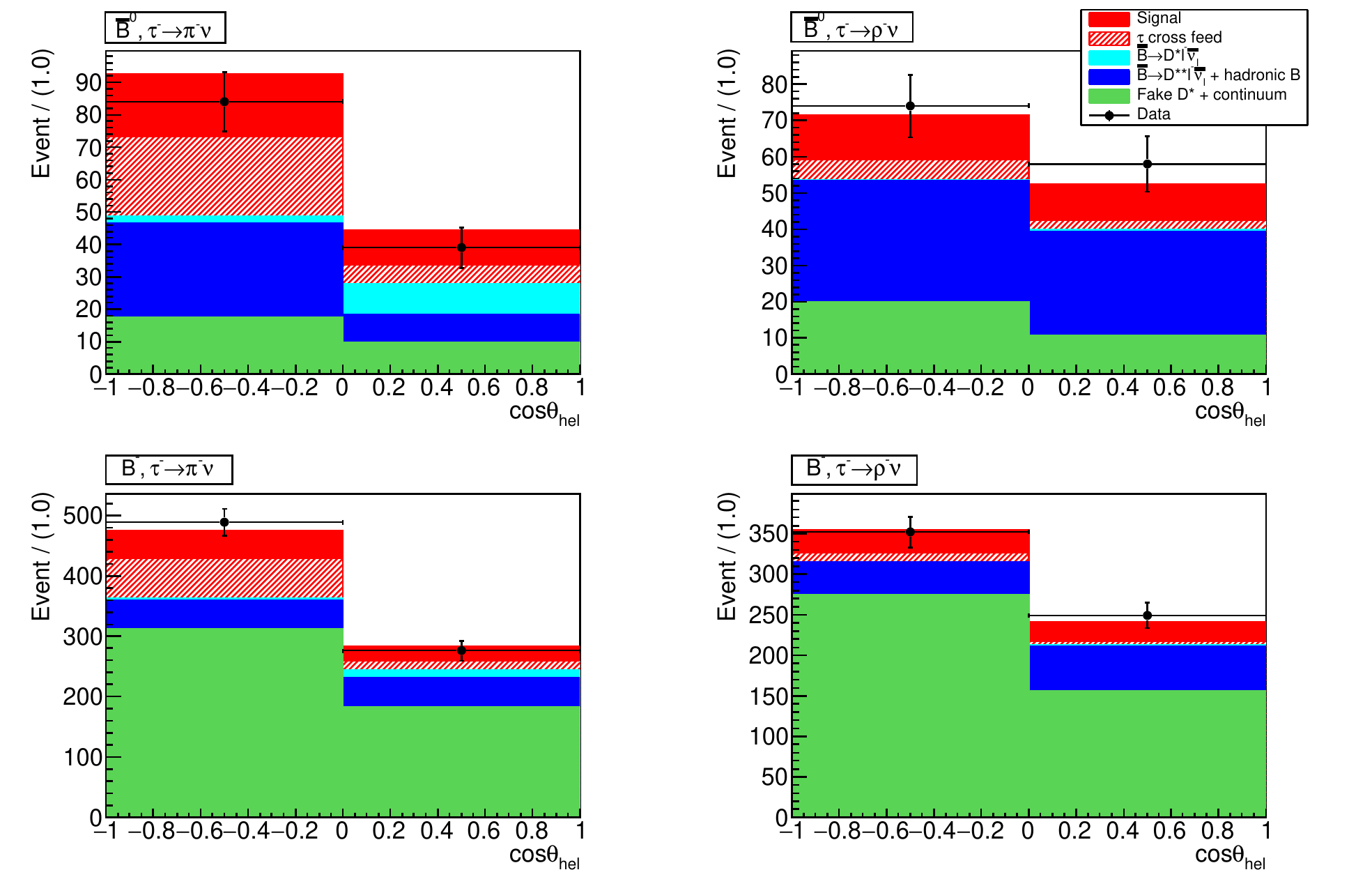}
  \caption{Fit result to the signal sample projected onto the $\cos\theta_{\rm hel}$ axis.}
  \label{fig-fitcoshel}
\end{figure}

\section{Result}

Figure~\ref{fig-fitresult} shows the fits to the signal and the normalization samples. (The figures in the forward and backward regions are shown in the Appendix~\ref{app-fwdbwd}.) The $\cos\theta_{\rm hel}$ distribution is shown in Fig.~\ref{fig-fitcoshel}. The observed signal and normalization yields are summarized in Table~\ref{tab-obtained-yield}. The $p$-values are found to be 15\% for the normalization fit and 29\% for the signal fit. From the fit, we obtain
\begin{eqnarray}
  R(D^*)   &=& 0.276 \pm 0.034({\rm stat.}) ^{+0.029} _{-0.026} ({\rm syst.}),\\
  P_{\tau} &=& -0.44 \pm 0.47({\rm stat.}) ^{+0.20} _{-0.17} ({\rm syst.}).
\end{eqnarray}
The signal significance is 9.7$\sigma$ (statistical error only) or 7.1$\sigma$ (including the systematic uncertainty). The significance is taken from $\sqrt{2 \ln (L_{\rm max} / {L_0})}$, where $L_{\rm max}$ and $L_0$ are the likelihood with the nominal fit and the null hypothesis, respectively.

\begin{table}[t!]
  \centering
  \caption{Observed signal and normalization yields. The errors are the statistical errors of the yields. As we use the common $R(D^*)$ and $P_{\tau}$ for all the samples as the fit parameters, these errors are correlated.}
  \begin{tabular}{@{\hspace{0.5cm}}l@{\hspace{0.5cm}}|@{\hspace{0.5cm}}c@{\hspace{0.5cm}}@{\hspace{0.5cm}}c@{\hspace{0.5cm}}}
    \hline
    \hline
    Sample & \multicolumn{2}{c}{Observed number of events}\\
    \hline
    \multicolumn{3}{c}{Signal}\\
    \hline
     & (Signal) & ($\tau$ cross feed)\\
    $({\bar B}^0, \pi^- \nu_{\tau})$  & $68.1 \pm 8.5$ & $82   \pm 10$\\
    $({\bar B}^0, \rho^- \nu_{\tau})$ & $51.1 \pm 6.4$ & $17.0 \pm  2.1$\\
    $(B^-, \pi^- \nu_{\tau})$         & $29.7 \pm 3.7$ & $30.8 \pm  3.8$\\
    $(B^-, \rho^- \nu_{\tau})$        & $21.9 \pm 2.7$ & $8.3  \pm  1.0$\\
    \hline
    \multicolumn{3}{c}{Normalization}\\
    \hline
    ${\bar B^0}$                      & \multicolumn{2}{c}{ $2504 \pm 52$ }\\
    $B^-$                             & \multicolumn{2}{c}{ $4699 \pm 81$ }\\
    \hline
    \hline
  \end{tabular}
  \label{tab-obtained-yield}
\end{table}

\begin{figure}[t!]
  \centering
 \includegraphics[width=11cm]{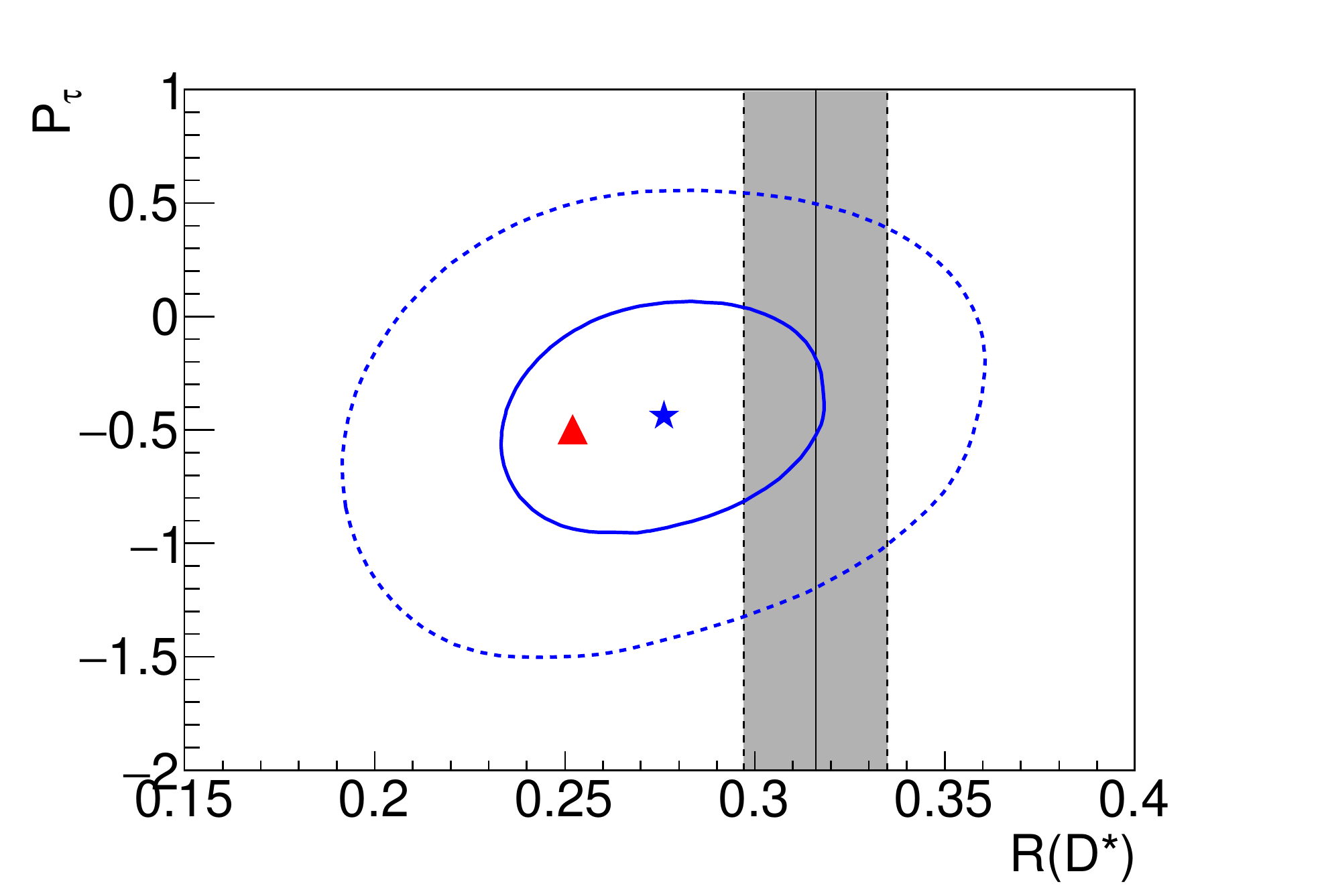}
  \caption{Comparison of our result (blue star for the best-fit value, contours with the blue solid and dashed lines for $1\sigma$ and $2\sigma$, respectively) with the SM prediction (red triangle). The gray region shows the average of the experimental results as of March 2016~\cite{cite-HFAG}.}
  \label{fig-significance}
\end{figure}

Figure~\ref{fig-significance} shows a comparison of our result with the theoretical prediction based on the SM~\cite{cite-RDst,cite-TW} in the $R(D^*) - P_{\tau}$ plane. The consistency of our result with the SM is $0.6\sigma$.

\section{Conclusion}

We report the measurement of $R(D^*)$ with hadronic $\tau$ decay modes $\tau^- \rightarrow \pi^- \nu_{\tau}$ and $\tau^- \rightarrow \rho^- \nu_{\tau}$ and the first measurement of $P_{\tau}$ in the decay ${\bar B} \rightarrow D^* \tau^- {\bar \nu_{\tau}}$, using $772 \times 10^6$ $B\bar{B}$ data accumulated with the Belle detector. Our preliminary results are
\begin{eqnarray}
  R(D^*)   &=& 0.276 \pm 0.034({\rm stat.}) ^{+0.029} _{-0.026} ({\rm syst.}),\\
  P_{\tau} &=& -0.44 \pm 0.47({\rm stat.}) ^{+0.20} _{-0.17} ({\rm syst.}),
\end{eqnarray}
which is consistent with the SM prediction within 0.6$\sigma$.

\begin{acknowledgements}

We thank the KEKB group for the excellent operation of the
accelerator; the KEK cryogenics group for the efficient
operation of the solenoid; and the KEK computer group,
the National Institute of Informatics, and the 
PNNL/EMSL computing group for valuable computing
and SINET4 network support.  We acknowledge support from
the Ministry of Education, Culture, Sports, Science, and
Technology (MEXT) of Japan, the Japan Society for the 
Promotion of Science (JSPS), and the Tau-Lepton Physics 
Research Center of Nagoya University; 
the Australian Research Council;
Austrian Science Fund under Grant No.~P 22742-N16 and P 26794-N20;
the National Natural Science Foundation of China under Contracts 
No.~10575109, No.~10775142, No.~10875115, No.~11175187, No.~11475187
and No.~11575017;
the Chinese Academy of Science Center for Excellence in Particle Physics; 
the Ministry of Education, Youth and Sports of the Czech
Republic under Contract No.~LG14034;
the Carl Zeiss Foundation, the Deutsche Forschungsgemeinschaft, the
Excellence Cluster Universe, and the VolkswagenStiftung;
the Department of Science and Technology of India; 
the Istituto Nazionale di Fisica Nucleare of Italy; 
the WCU program of the Ministry of Education, National Research Foundation (NRF) 
of Korea Grants No.~2011-0029457,  No.~2012-0008143,  
No.~2012R1A1A2008330, No.~2013R1A1A3007772, No.~2014R1A2A2A01005286, 
No.~2014R1A2A2A01002734, No.~2015R1A2A2A01003280 , No. 2015H1A2A1033649;
the Basic Research Lab program under NRF Grant No.~KRF-2011-0020333,
Center for Korean J-PARC Users, No.~NRF-2013K1A3A7A06056592; 
the Brain Korea 21-Plus program and Radiation Science Research Institute;
the Polish Ministry of Science and Higher Education and 
the National Science Center;
the Ministry of Education and Science of the Russian Federation and
the Russian Foundation for Basic Research;
the Slovenian Research Agency;
Ikerbasque, Basque Foundation for Science and
the Euskal Herriko Unibertsitatea (UPV/EHU) under program UFI 11/55 (Spain);
the Swiss National Science Foundation; 
the Ministry of Education and the Ministry of Science and Technology of Taiwan;
and the U.S.\ Department of Energy and the National Science Foundation.
This work is supported by a Grant-in-Aid from MEXT for Science Research 
in a Priority Area (``New Development of Flavor Physics''), 
from JSPS for Creative Scientific Research (``Evolution of Tau-lepton Physics''), 
a Grant-in-Aid for Scientific Research (S) “Probing New Physics with Tau-Lepton” (No. 26220706) 
and was partly supported by a Grant-in-Aid for JSPS Fellows (No. 25.3096).
We thank Y.~Sakaki, R.~Watanabe and M.~Tanaka for their invaluable suggestions and helps.

\end{acknowledgements}

\appendix*

\section{Additional figures of the fit result}\label{app-fwdbwd}

\begin{figure}[t!]
  \centering
  \includegraphics[width=15cm]{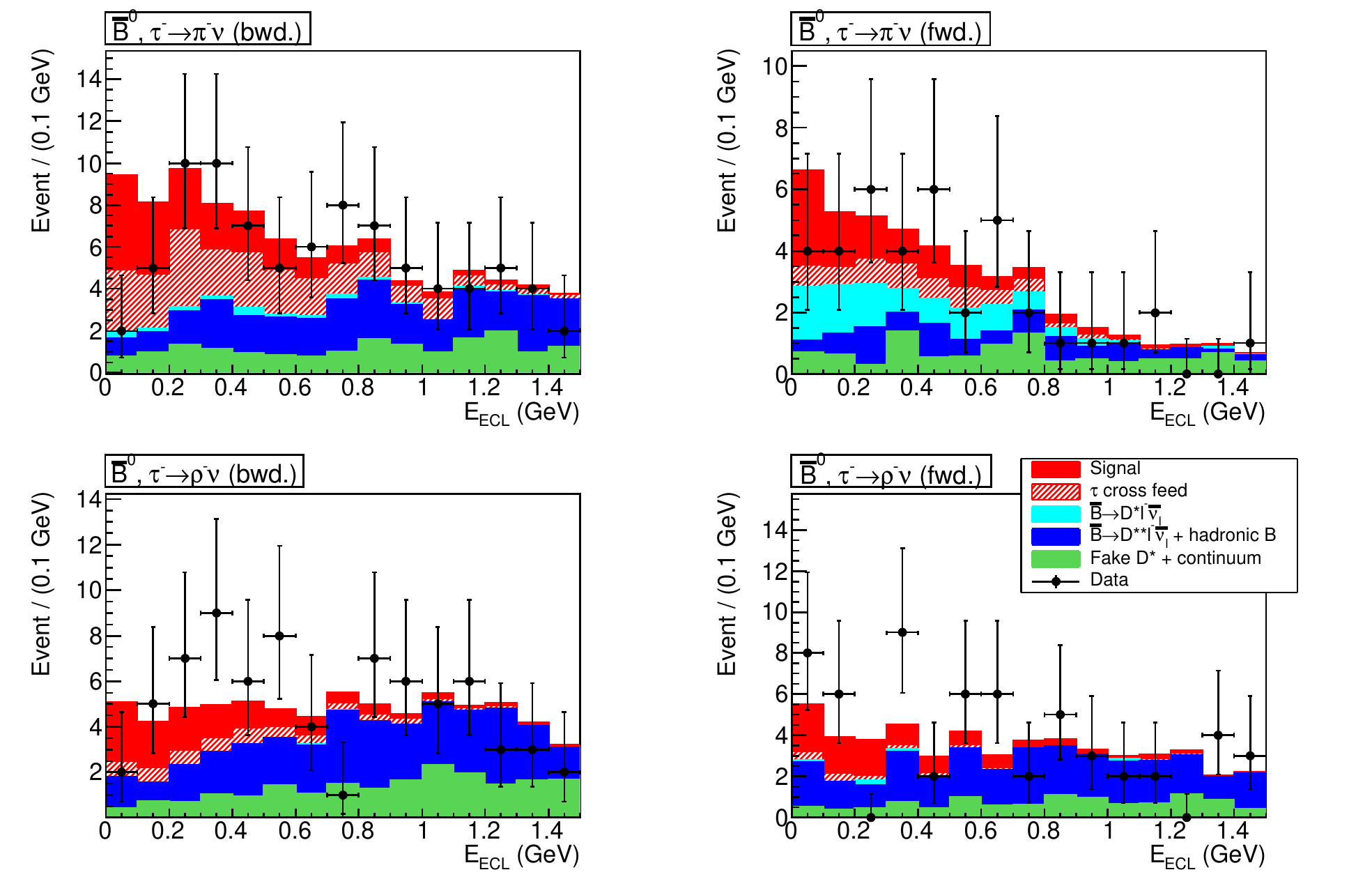}
  \includegraphics[width=15cm]{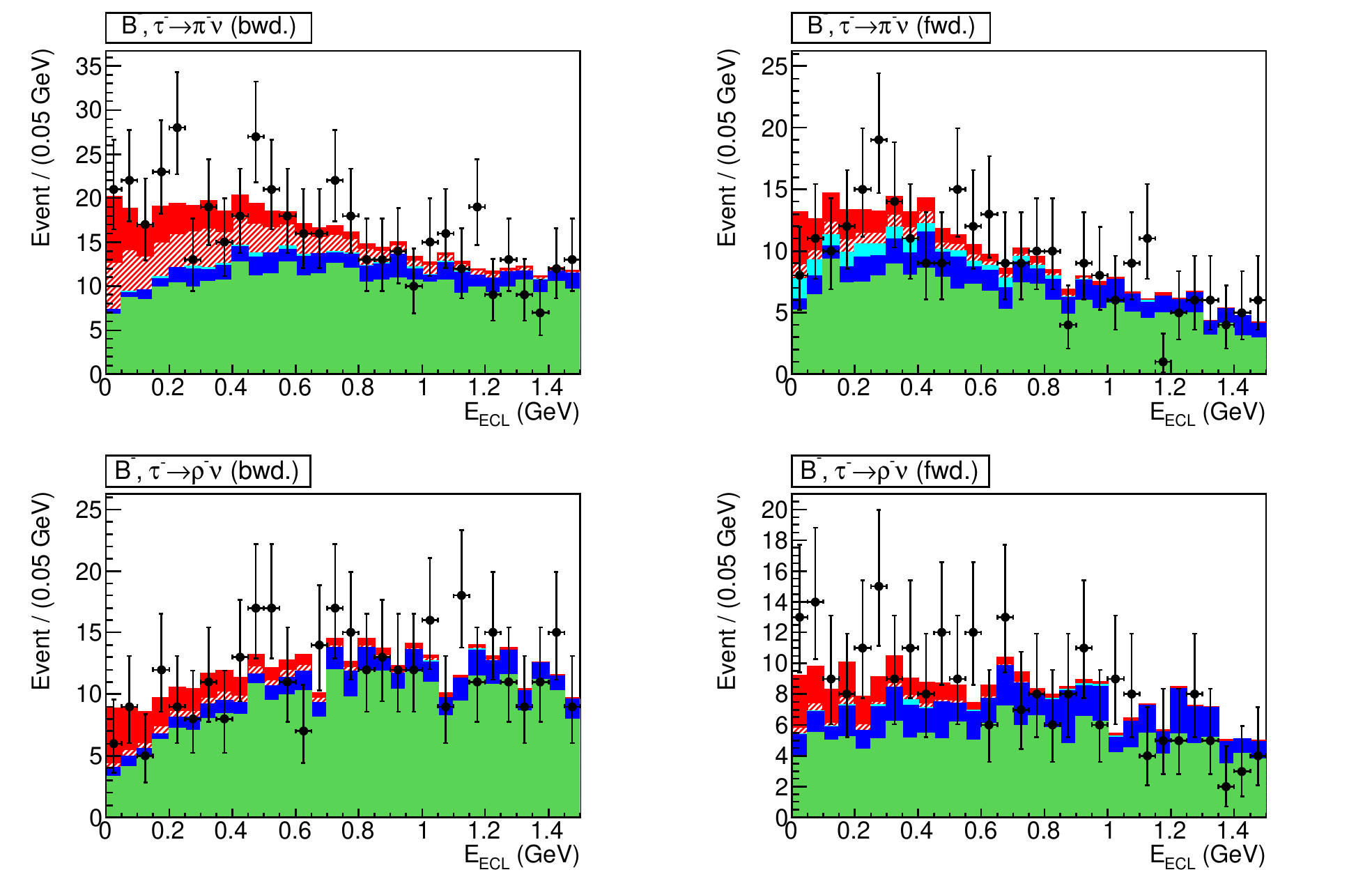}
  \caption{Fit results to the four signal samples. The left and right columns are for the backward and the forward bins of $\cos\theta_{\rm hel}$, respectively.}
  \label{fig-app-fit-separate}
\end{figure}

Figure~\ref{fig-app-fit-separate} shows the fit results in the forward and backward regions.


\begin{thebibliography}{99}
\bibitem{cite-CC}
  Throughout this paper, the inclusion of the charge conjugate mode decay is implied.
\end{thebibliography}
\end{document}